\documentclass[12pt]{article} 
\usepackage{jheppub}

\usepackage{amsmath,amsfonts,amsbsy,amssymb,array,accents,dsfont}
\usepackage{enumerate}
\usepackage[shortlabels]{enumitem}
\usepackage{graphicx} 
\usepackage{caption} 
\usepackage{subcaption} 
\usepackage{mathrsfs}
\usepackage{upgreek}
\usepackage{bm}
\usepackage{slashed}

\usepackage{xparse}
\usepackage{xspace}

\NewDocumentCommand\eqn{om}{%
  \IfNoValueTF{#1}
     {\[ #2 \]}
     {\begin{equation}\label{#1} #2  \end{equation} \expandafter\newcommand\csname #1\endcsname{\eqref{#1}\xspace}\ignorespaces}
}
\NewDocumentCommand\eqna{om}{%
  \IfNoValueTF{#1}
    {\begin{align*} #2 \end{align*}}
    {\begin{equation}\label{#1}\begin{split} #2  \end{split}\end{equation} \expandafter\def\csname #1\endcsname{\eqref{#1}\xspace}\ignorespaces}
}

\newcommand{\rcite}{\cite}

\def\alphab{{\boldsymbol\alpha}}

\def\lambdab{{\boldsymbol\lambda}}

\def\xx{{\bf  x}}

\def\sl{\text{sl}}
\def\su{\text{su}}

\def\eps{\epsilon}
\def\vareps{\varepsilon}

\def\sltwo{\ensuremath{SL(2,\bR)}}

\def\sutwo{{SU(2)}}

\def\tight#1{\! #1 \!}  

\def\({\left(}
\def\){\right)}
\def\[{\left[}
\def\]{\right]}

\def\ie{{i.e.}}
\def\eg{{e.g.}}

\def\etc{{etc}}

\def\lstr{\ell_{\textit{s}}}

\def\nfive{{n_5}}

\def\sfA{{\mathsf A}}		\def\sfB{{\mathsf B}}				
		\def\sfF{{\mathsf F}}		\def\sfG{{\mathsf G}}		
		\def\sfJ{{\mathsf J}}				\def\sfL{{\mathsf L}}
						
\def\sfQ{{\mathsf Q}}						
					\def\sfX{{\mathsf X}}
		\def\sfZ{{\mathsf Z}}
		\def\sfb{{\mathsf b}}		\def\sfc{{\mathsf c}}

\def\sfm{{\mathsf m}}		\def\sfn{{\mathsf n}}				
\def\sfq{{\mathsf q}}						
		\def\sfv{{\mathsf v}}				\def\sfx{{\mathsf x}}
\def\sfy{{\mathsf y}}

\DeclareMathSymbol{\medhatsym}{\mathord}{largesymbols}{"62} 

\DeclareMathSymbol{\medtildesym}{\mathord}{largesymbols}{"65}

\def\Ry{R_y}

\newcommand{\rhoo}{\ensuremath{\! \rho \:\! }}

\def\alphab{{\boldsymbol\nu}}

\def\half{\frac12}

\def\hf{\coeff12}
\def\tr{{\rm Tr}}
\def\One{{\hbox{1\kern-1mm l}}}

\def\barray{\begin{array}}
\def\earray{\end{array}}
\def\be{\begin{equation}}
\def\ee{\end{equation}}
\def\bea{\begin{align}}
\def\eea{\end{align}}
\def\nn{\nonumber}

\newcommand{\bR}{{\mathbb R}}
\newcommand{\bS}{{\mathbb S}}
\newcommand{\bT}{{\mathbb T}}

\newcommand{\bZ}{{\mathbb Z}}

\def\bbT{\mathbb{T}}

\def\R{R_y}

\definecolor{cardinal}{rgb}{0.6,0,0}
\definecolor{darkgreen}{rgb}{0,0.4,0}
\definecolor{green}{rgb}{0,0.4,0}
\definecolor{golden}{rgb}{0.92, 0.7, 0}
\definecolor{midnight}{rgb}{0, 0, 0.5}
\definecolor{darkblue}{rgb}{0, 0, 0.7}

\numberwithin{equation}{section}

\mathchardef\mhyphen="2D

\def\cD{\mathcal {D}}  \def\cF{\mathcal {F}}
\def\cG{\mathcal {G}} \def\cH{\mathcal {H}} 
\def\cJ{\mathcal {J}}  
\def\cM{\mathcal {M}} \def\cN{\mathcal {N}} 
\def\cP{\mathcal {P}} \def\cQ{\mathcal {Q}} 
\def\cS{\mathcal {S}}  
\def\cV{\mathcal {V}} \def\cW{\mathcal {W}} \def\cX{\mathcal {X}}
\def\cY{\mathcal {Y}} \def\cZ{\mathcal {Z}}

\def\one{{\hbox{\kern+.5mm 1\kern-.8mm l}}}
\def\zero{{\hbox{0\kern-1.5mm 0}}}

\newcommand{\ket}[1]{{\,| {#1} \rangle}}

\newcommand{\T}[3]{\ensuremath{ #1{}^{#2}_{\phantom{#2} \! #3}}}		

\def\id{\textrm{id}}

\def\id{{1 \kern-.28em {\rm l}}}

\def\journal#1&#2(#3){\unskip, \sl #1\ \bf #2 \rm(19#3) }
\def\andjournal#1&#2(#3){\sl #1~\bf #2 \rm (19#3) }

\def\ie{{\it i.e.}}
\def\eg{{\it e.g.}}

\def\etc{{\it etc}}

\def\sst{\scriptscriptstyle}

\def\half{\frac12}
\def\hf{{\textstyle\half}}
\def\One{{1\hskip -3pt {\rm l}}}

\catcode`\@=11
\def\slash#1{\mathord{\mathpalette\c@ncel{#1}}}
\def\eps{\epsilon}
\def\vareps{\varepsilon}
\def\underrel#1\over#2{\mathrel{\mathop{\kern\z@#1}\limits_{#2}}}

\catcode`\@=12

\def\ket#1{\left| #1\right\rangle}
\def\vev#1{\left\langle #1 \right\rangle}

\def\tr{{\rm tr}}

\def\exp{{\rm exp}}

\def\ie{{\it i.e.}}
\def\eg{{\it e.g.}}

\def\mbar{{\bar m}}

\title{
On the BPS sector in \texorpdfstring{$\bf AdS_3/CFT_2$}{} Holography
}

\author{Emil J. Martinec$^a$, Stefano Massai$^{b,c}$ {\it and}\,
  David Turton$^d$\\}

\affiliation[a]{
Kadanoff Center for Theoretical Physics and Enrico Fermi Institute\\ 
University of Chicago\\ 
5640 S. Ellis Ave.\\
Chicago IL 60637\\ 
}

\affiliation[b]{
Dipartimento di Fisica e Astronomia ``Galileo Galilei''\\
Universit\`a di Padova, Via Marzolo 8, 35131 Padova, Italy\\
}

\affiliation[c]{
INFN, Sezione di Padova, Via Marzolo 8, 35131 Padova, Italy\\
}

\affiliation[d]{
Mathematical Sciences and STAG Research Centre, University of Southampton, \\
Highfield, Southampton, SO17 1BJ, UK\\
}

 \emailAdd{ejmartin@uchicago.edu}
 \emailAdd{stefano.massai@pd.infn.it}
 \emailAdd{d.j.turton@soton.ac.uk}

\abstract{ 
The BPS sector in $AdS_3/CFT_2$ duality has been fertile ground for the exploration of gauge/gravity duality, from the match between black hole entropy and the CFT elliptic genus to the construction of large families of geometrical microstates and the identification of the corresponding states in the CFT.  
Worldsheet methods provide a tool to further explore the relation between string theory in the bulk and corresponding CFT quantities.  We show how to match individual BPS strings to their counterparts in the symmetric product orbifold CFT.  In the process, we find an exact match between known constructions of microstate geometries and condensates of BPS supergraviton strings, and discuss their role in the broader collection of BPS states.  In particular, we explore how microstate geometries develop singularities; and how string theory resolves these singularities through the appearance of ``tensionless'' string dynamics, which is the continuation of structures found in the weak-coupling CFT into the strongly coupled regime described by string theory in the bulk.  We argue that such ``tensionless'' strings are responsible for black hole microstructure in the bulk description.
}

\begin{document}
\hypersetup{pageanchor=false}
\begin{titlepage}
\maketitle
\thispagestyle{empty}
\end{titlepage}
\hypersetup{pageanchor=true}
\pagenumbering{arabic}






\section{Introduction and Summary} 
\label{sec:intro}

The $AdS_3/CFT_2$ duality of the string theory onebrane-fivebrane system exhibits an extensively developed holographic dictionary (reviews include~\rcite{David:2002wn,Skenderis:2008qn,Shigemori:2020yuo,Bena:2022ldq}).  In particular, there is a rich variety of examples of CFT microstates that have been matched to corresponding smooth horizonless geometries with $AdS_3\times\bS^3\times\cM$ asymptotics, where $\cM=\bT^4$ or $K3$ (see~\rcite{Shigemori:2020yuo,Bena:2022ldq} for recent overviews and further references).

In early work, the 1/2-BPS ground states of the system were studied.%
\footnote{Our normalization of the fraction of supersymmetry preserved is relative to the $\sltwo$ invariant NS-NS vacuum state.  Thus Ramond-Ramond ground states are 1/2-BPS, and breaking the left-moving half of the remaining supersymmetry via chiral momentum excitations results in 1/4-BPS states.}
The different brane bound state configurations source supergravity solutions known as {\it supertubes}~\rcite{Lunin:2001fv,Lunin:2001jy,Taylor:2005db,Kanitscheider:2007wq}.  There is an explicit map between CFT ground states and supergravity solutions, which we review in section~\ref{sec:supertubes} below.  The entropy of these ground states is given in terms of the charge quanta $n_1,n_5$ of onebranes and fivebranes, as well as angular momentum $J_L$ on $\bS^3$, by~\rcite{Lunin:2002qf,Palmer:2004gu}
\be
\label{halfBPSent}
S_{\half\textrm{-BPS}}=2\pi\sqrt{\frac c6\big(n_5 n_1 - |J_L|\big)}  
\ee
where $c=12$ for $\bT^4$, and $c=24$ for $K3$.

Exciting the system away from any of these ground states by adding momentum-carrying supergravity waves leads to a collection of horizonless 1/4-BPS NS5-F1-P geometries known as {\it superstrata}~\rcite{Bena:2015bea,Bena:2016ypk,Bena:2017xbt,Ceplak:2018pws,Heidmann:2019zws,Ceplak:2022wri} (for a review, see~\rcite{Shigemori:2020yuo}). 
Each geometry has a well-understood holographic map to a (coherent) state in the symmetric product CFT; this map is reviewed and elaborated upon in section~\ref{sec:superstrata} below.%
\footnote{As we review in section~\ref{sec:symprod}, the symmetric product CFT and the bulk effective string theory occupy complementary regions of the moduli space; however, BPS protected quantities such as index states are robust and can thus be compared.  
In addition to superstrata, multi-centered, bubbled geometries~\rcite{Bena:2005va} have also been constructed, which are BPS for particular values of the moduli but lift off the BPS bound as the moduli are deformed to generic values~\rcite{Bossard:2019ajg}.  Because they are only accidentally BPS at a particular locus in the moduli space, these bubbled geometries are not protected and it is not guaranteed that they should map to some well-defined collection of states in the symmetric product CFT.
} 

Both supertubes and superstrata are examples of {\it smooth, horizonless BPS microstate geometries}.
The matching of 1/2-BPS states was the first example where the microstates in the CFT were mapped one-for-one onto fully back-reacted supergravity geometries.  One can think of the superstratum construction as extending this holographic map to a much larger class of 1/4-BPS CFT states for which the fully back-reacted supergravity solution is known.

Furthermore, an analysis~\rcite{deBoer:1998us} of the elliptic genus,
\be
Z_{\rm\sst EG}(q,y) = \tr\Big[(-1)^{F_L+F_R}\,q^{n_p-\frac{c}{24}}\, y^{J_0}  \Big]  ~,
\ee
exhibits an exact match between the spacetime CFT and the corresponding index of BPS states in a gas of supergravitons, up to level $n_p=\frac14 n_1n_5$.  Superstrata provide a fully back-reacted bulk realization of a coherent state basis of this supergraviton gas, which includes the index states.

However, the superstratum construction is not restricted to the regime $n_p< \frac14 n_1n_5$.  The quantum numbers can be extended far into the regime where the number of momentum quanta $n_p$ is much larger than either $n_1n_5$ or $J_L$, and in particular into the regime $n_1n_5n_p>J_L^2$.  In this regime, the generic element of the density of states is a BTZ$\times \mathbb{S}^3$ black string in 6d, which reduces to a BMPV~\rcite{Breckenridge:1996is} black hole in 5d.
The entropy of superstratum states in this regime has been estimated to be~\rcite{Shigemori:2019orj}
\be
\label{stratument}
S_{\rm geom} \sim \sqrt{n_5 n_1} \, n_p^{1/4} ~,
\ee
which for generic large charges is much smaller than the 1/4-BPS BTZ black hole entropy,
\be
\label{BHent}
S_{\rm BTZ} = 2\pi\sqrt{n_5 n_1 n_p - J_L^2} ~.
\ee
In the BTZ regime $n_1n_5n_p>J_L^2$, the CFT elliptic genus has the same asymptotic growth as the black hole density of states (thus, almost all 1/4-BPS BTZ black holes are bosons).

The fact that the number of smooth supergravity geometries is subleading in the entropy answers a question that has sometimes been asked, namely whether to include the black hole geometry in the sum over saddles in the Euclidean gravitational path integral, if one is proposing to replace the black hole by an ensemble of stringy ``fuzzball'' states.
The answer is yes, it should be included~-- the Euclidean black hole solution represents all the generic members of the ensemble that can't be distinguished at the level of supergravity.  The geometrical microstates represent auxiliary complex saddles in the Euclidean path integral, and don't result in overcounting since they represent a vastly subleading contribution.  The use of the black hole solution does not necessarily mean that the microstates so represented lack the horizon-scale structure posited by the fuzzball proposal; it simply means that, as seen from outside by supergravity probes, any such structure will appear to be well-approximated by the black hole solution and its properties.

The 1/4-BPS black hole solution accounts for the bulk of the 1/4-BPS entropy.  As yet, we don't have a bulk picture of the microstates, or a detailed understanding of the holographic map.  These states appear to have a horizon in their effective supergravity description; however, the fuzzball proposal posits that all microstates are fundamentally horizonless.  This issue is perhaps the central open question of the fuzzball program~-- whether the internal degrees of freedom of any microstate are in causal contact with the exterior spacetime.

\begin{figure}[ht]
\centering
\includegraphics[width=0.6\textwidth]{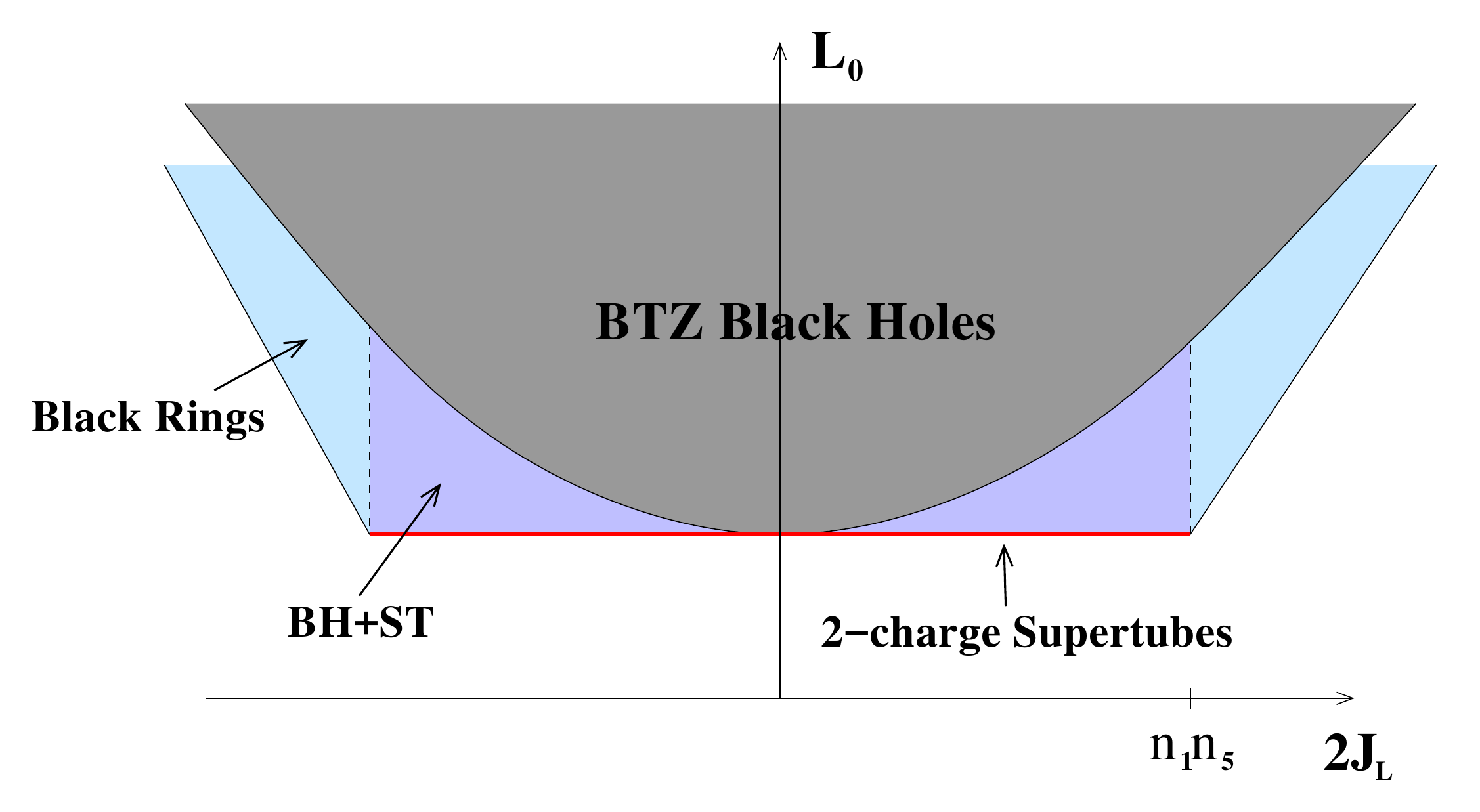}
\caption{\it Additional BPS black objects inhabit the region between the BTZ threshold and the BPS ground states.    }
\label{fig:enigmatic}
\end{figure}

While BTZ black holes are the generic elements of the ensemble of states above the BTZ threshold $n_5n_1n_p=J_L^2$, below this threshold there are other 1/4-BPS black objects~\rcite{Bena:2011zw} which dominate the density of states.  Depending on the regime of parameters, the dominant configuration is either a zero-angular momentum black hole surrounded by a supertube that carries the angular momentum, or a black ring; see figure~\ref{fig:enigmatic}.  The corresponding entropies are
\begin{align}
\label{enigma-ent}
\begin{split}
S_{\sst\rm BH+ST} &= 2\pi \sqrt{n_5n_1n_p}\,\bigg(1-\sqrt{\frac{2J_L-n_p}{n_1n_5}} \,\bigg)
~~~~\quad\qquad\qquad\qquad,\qquad~~ 0<J_L<\frac{n_1n_5}2~,
\\[.3cm]
S_{\rm ring} &= 2\pi\sqrt{n_5n_1(n_5n_1+n_p-2J_L)}\,\bigg(1-\sqrt{\frac{n_5n_1-n_p}{n_1n_5}} \,\bigg)
\quad,\quad \frac{n_1n_5}2 < J_L < n_1n_5~.
\end{split}
\end{align}
Initially, the portions of the phase diagram below the BTZ threshold were not well understood, and so they became known as {\it enigmatic phases}.
%
%
There are also smooth microstate geometries in these sub-BTZ regimes; once again their entropy is generically subleading, \eg\ for $J_L \ll n_1n_5/2$ one has~\rcite{Shigemori:2019orj} 
\be
\label{light-ent}
S_{\rm geom} \sim (n_5 n_1)^{1/4} n_p^{1/2}
~~,~~~~
n_p \ll n_1n_5  ~.
\ee

As mentioned above, the elliptic genus in the regime near the maximally spinning ground state%
\footnote{This regime is the spectral flow of the NS-NS sector states with $n_p\tight< \frac{n_1n_5}{4}$ to the R-R sector.} ($\frac{n_1n_5}{4}\tight<n_p\tight< J$, with $\frac{n_1n_5}4\tight<J\tight<\frac{3n_1n_5}{4}$) is accounted for by smooth supergravity solutions.  The growth of index states
is even slower than the growth of the 1/4-BPS supergravity partition function~\eqref{light-ent}, let alone that of the relevant black object~\eqref{enigma-ent};
for instance, along the line $J=\half n_1n_5$, one has the density of index states (computed either in the CFT or in supergravity)~\rcite{Benjamin:2015vkc,Benjamin:2016pil}
\be
\label{light-EG}
\rho^{\rm\sst EG}_{\rm S^N(K3)}(n_p) \sim N\,\exp\Big(2\pi\sqrt{12n_p} \Big)  ~.
\ee
As a consequence, none of the 1/4-BPS black holes dominating the enigmatic phases are index states, and might not lie exactly on the BPS bound; for instance, they could be lifted slightly away from the BPS bound by perturbative corrections to the classical solution.  Regardless, there are exponentially many more of them than the index states, with the latter being entirely accounted for by the supergravity solutions in this triangular region.  In fact, we see from~\eqref{light-ent} that there are also exponentially more 1/4-BPS supergravity states than there are index states.%
\footnote{Indeed, recent analysis of the $AdS_2$ near-horizon region of the BTZ solution~\rcite{Heydeman:2020hhw} suggests that the index states are separated by a gap of order $1/(n_1n_5)$ from a quasi-continuum of slightly non-BPS states.  Naively, this gap prevents unprotected states from lying exactly on the BPS bound, otherwise they could not lift smoothly as one deforms the CFT moduli. It would be interesting to see whether indeed perturbative corrections to classically BPS states lift them slightly off the BPS bound.}

Given that the geometrical index states are interspersed with a large collection of unprotected black hole states at and near the BPS bound, one expects that a slight perturbation of the former will thermalize into the latter.  This feature provides a key rationale for the exploration of horizonless geometries~-- they lie deep within the black hole phase, while being amenable to analysis in the bulk description.  They thus provide a portal to the regime of generic black hole states, and we can ask what processes are at work as they access this regime.  
We will explore in section~\ref{sec:singularities} a particular mechanism, in which singularities which can develop in the bulk geometry signal the onset of a deconfinement transition in the underlying fivebrane dynamics.  The deconfined phase is expected to describe generic black hole states.

In this work, we review and expand upon the above detailed picture of the BPS sector in $AdS_3/CFT_2$ holography, concentrating on the enigmatic regime below the BTZ threshold.  The zoology of these states can be organized into three general classes.  Smooth geometrical microstates are the least entropic, while black holes are the most entropic; in between one has what one might call
{\it perturbative stringy horizonless microstates}.  These consist of a BPS gas of perturbative strings placed in any microstate geometry.  The back-reaction of these string sources leads to classical singularities that are resolved by perturbative string effects.  The entropy of such strings exceeds the supergravity entropy~\eqref{light-ent}, but again falls short of the black object entropy~\eqref{enigma-ent}.  We estimate this entropy in section~\ref{sec:stringyBPS} and find a Hagedorn spectrum of 1/4-BPS string states; in particular, near $J_L=n_1n_5/2$, we find
\be
\label{Hag spec}
S_{\rm pert} = 2\pi\sqrt{\frac{2}{n_5}n_p\big(n_p  +n_1n_5-2J_L \big)} ~,
\ee
which indeed lies between~\eqref{light-ent} and~\eqref{enigma-ent}.

Our analysis employs the methods of worldsheet string theory to explore these different families of BPS states, and the connections between them.  
The main tool in this analysis is the worldsheet construction of supersymmetric ground states developed in%
~\rcite{Martinec:2017ztd,Martinec:2018nco,Martinec:2019wzw,Martinec:2020gkv,Bufalini:2021ndn,Bufalini:2022wyp,Bufalini:2022wzu}.
After reviewing relevant aspects of the spacetime CFT in section~\ref{sec:symprod}, we provide an overview of the gauged WZW model on the worldsheet that describes a family of heavy BPS ground states in section~\ref{sec:setup}.  

As a warmup exercise, we review in section~\ref{sec:hfBPS} the construction in~\rcite{Martinec:2020gkv} of worldsheet vertex operators that describe 1/2-BPS deformations of 1/2-BPS supertube backgrounds.  
The BPS vertex operator spectrum of the worldsheet theory mediates transitions between BPS states, perturbatively around a given BPS state.  
Condensation of 1/2-BPS vertex operators (\ie\ exponentiating them into the worldsheet action) allows us to explore the nearby 1/2-BPS configuration space of supertubes, out to some finite distance. 

The family of 1/2-BPS NS5-F1 backgrounds is well-understood at the level of bulk supergravity~\rcite{Lunin:2001fv,Lunin:2001jy,Taylor:2005db,Kanitscheider:2007wq}.  One can think of them as being specified in part by the shape of the fivebranes that results from their back-reaction on the condensate of fundamental strings they are carrying.
We show how the worldsheet theory exhibits the stringier aspects of these configurations, such as how the worldsheet theory codes the source profile and is thus able to determine the location of the fivebranes in the background.

We then extend the analysis to 1/4-BPS excitations, which add momentum charge to the system.  In section~\ref{sec:superstrata} we consider the 1/4-BPS worldsheet vertex operators that describe deformations within supergravity.  These operators have (say) BPS polarization states on the right-movers and arbitrary polarization on the left-moving side. We elaborate a precise map between the 1/4-BPS vertex operators that describe supergravity modes and the known methods of constructing superstrata.  

Condensing these 1/4-BPS excitations into the background generates the smooth NS5-F1-P superstratum geometries.  We exhibit a one-to-one correspondence between the 1/4-BPS spectrum and the superstratum modes that have been studied in the literature, and thus provide evidence that all the smooth 1/4-BPS geometries in a finite neighborhood of the initial 1/2-BPS background have in principle been found.  As we noted above, the result of~\rcite{deBoer:1998us} shows that these geometries saturate the elliptic genus up to level $\frac14 n_1n_5$, and thus provide the explicit bulk geometries that contribute to this supersymmetric index in the sub-BTZ regime where~\eqref{light-EG} holds.

There are also 1/4-BPS vertex operators describing excited string states.  
In section~\ref{sec:stringyBPS}, we consider these 1/4-BPS perturbative string excitations that lie outside of supergravity.  These are described by worldsheet vertex operators that are in BPS ground states for say the worldsheet right-movers but have arbitrary oscillator excitation for the left-movers, subject to the BRST constraints.  The backgrounds sourced by BPS ensembles of such strings fall into an intermediate category, in between that of microstate geometries and that of generic fuzzballs, that we  call ``perturbatively stringy horizonless microstates''.  We will show that the number of perturbatively stringy microstates has the Hagedorn entropy~\eqref{Hag spec}, and so is parametrically larger than that of microstate geometries~\eqref{light-ent} while still falling short of that of generic fuzzballs~\eqref{enigma-ent}.

These perturbative stringy microstates are unable to realize the maximal degree of fractionation of momentum carriers seen in the weakly coupled CFT.  It is these highly fractionated momentum carriers that are responsible for the BTZ entropy in the weakly coupled CFT.  The lower degree of fractionation in the smooth geometries and the perturbative stringy microstates distinguishes them from generic microstates.

The worldsheet also allows us to see some of the stringy phenomena that occur as the geometrical approximation begins to break down.  In section~\ref{sec:singularities}, 
we consider potential singularities that might arise at the non-linear level when we deform away from the points in configuration space described by exactly solvable worldsheet theories.  One possibility, discussed in~\rcite{Eperon:2016cdd,Marolf:2016nwu}, is an ``instability'' in which excitations pile up at the locus of deepest redshift in the background.  We show how these analyses connect to the worldsheet theory, which resolves the singularities by showing, as suggested in~\rcite{Marolf:2016nwu}, that these lowest energy excitations are simply those which smoothly deform the initial 1/2-BPS background along the 1/2-BPS configuration space, and that having them pile up in the depths of the geometry and backreact is simply the mechanism by which they exponentiate into finite deformations of the background, rather than a singularity that signals the onset of the black hole phase.   

We also show how actual singularities can arise, when fivebranes in a supertube background self-intersect.  At the intersection locus, the supertube develops a vanishing two-cycle.  Wrapping D3-branes around this two-cycle leads to a ``tensionless'' string.%
\footnote{We put quotes around ``tensionless'' because it does not imply that there is no gap in the spectrum; see for instance~\rcite{Gaberdiel:2018rqv}.}
These strings are similar in many respects to the weak-coupling ``tensionless'' strings described by the symmetric orbifold.%
\footnote{A major theme running through the analysis is the close relation between individual BPS strings in the bulk description and cycles of the symmetric product CFT.  Of course, one is comparing states at vastly disparate points in the moduli space of the theory; it is the BPS property and associated non-renormalization theorems that permit a comparison of appropriate protected quantities.}

This ``tensionless'' string singularity signals the Hawing-Page phase transition in which the non-abelian degrees of freedom of the underlying fivebrane dynamics are liberated.  The regime of smooth, horizonless geometry is one where the fivebranes are slightly separated (as one sees for instance in the construction of 1/2-BPS backgrounds reviewed in section~\ref{sec:supertubes}), which abelianizes the fivebrane dynamics by giving mass to these effective strings.
The regime of smooth horizonless geometries thus appears to be intermingled with but distinct from the regime of generic microstates.

We will argue that the bulk of the entropy in the black hole regime comes from a gas of these effective strings, that arises as the geometrical approximation breaks down~\rcite{Martinec:2019wzw,Martinec:2020gkv}.  These strings appear to be the avatar of the entropic degrees of freedom of the symmetric product CFT in the regime of CFT strong coupling.

These light effective strings are the ``W-strings'' of {\it little string theory}, the strongly-coupled 6d self-dual string dynamics that governs a stack of decoupled fivebranes~\rcite{Dijkgraaf:1996cv,Maldacena:1996ya,Dijkgraaf:1997ku,Seiberg:1997zk} (for reviews, see~\rcite{Aharony:1999ks,Kutasov:2001uf}).  Thus the black hole phase transition in $AdS_3/CFT_2$ is conceptually no different than its cousins in $AdS_4$ and above, in which the black hole phase is associated to the deconfinement of non-abelian constituents of the underlying brane dynamics.  

We can see all this structure in bulk string theory on $AdS_3\times\bS^3\times\cM$ realized as the decoupling limit of the NS5-F1 system, because NS5-branes are solitonic objects of closed string dynamics.  Their tension scales as $1/g_s^2$, and so worldsheet string theory necessarily incorporates their back-reaction on geometry.%
\footnote{Different duality frames realize the light effective string differently.  In the NS5-F1 frame the little string is a fractional fundamental string; in the D5-D1 frame it is a fractional D1 realized as an instanton string in the D5 gauge theory.  Our ability to see the details of microstates varies from frame to frame; we choose the NS5-F1 frame precisely because stringy aspects of microstructure are more readily apparent.  It would be interesting to understand whether and how the mechanisms we discuss here are manifested in other duality frames.}
While it is often said that the background branes ``dissolve into flux'' in AdS/CFT, the worldsheet dynamics is smart enough to keep track of where the fivebranes are in the background (information that is non-perturbative in $\alpha'$), and to exhibit the mechanism of the deconfinement transition of the CFT, deep down at the bottom of the $AdS_3$ throat in the bulk description.


\section{The symmetric product orbifold and its BPS spectrum} 
\label{sec:symprod}

\subsection{Structure of the moduli space}
\label{sec:modspace}

We begin with a discussion of where the symmetric product orbifold lies in the moduli space of NS5-F1 backgrounds.
The NS5-F1 charge quanta $(n_5,n_1)$ are components of a charge vector $\bf q$ transforming in the $\underline{10}$ of the $O(5,5;\bZ)$ U-duality group of type II string theory on $\bT^4$.  
The CFT central charge $c=6N$ is a U-duality invariant written in terms of the symplectic inner product $N=\vev{{\bf q},{\bf q}}$.

The moduli space of the spacetime CFT has a number of weak-coupling cusps, one for each factorization of $N$ into a pair of integers $N=n_5n_1$. 
The background charge $\bf q$ breaks the U-duality symmetry down to the ``little group'' $\Gamma_{\bf q}$ that fixes the charge vector $\bf q$, which is a proper subgroup of the naive little group $O(5,4;\bZ)$.  The moduli space of the spacetime CFT is then 
\be
\frac{O(5,4;\bR)}{O(5,\bR)\times O(4,\bR)}\Big/\Gamma_{\bf q}  ~.
\ee
On the other hand, the vacuum moduli space of string theory on $\bT^4$ is
\be
\frac{O(5,5;\bR)}{O(5,\bR)\times O(5,\bR)}\Big/O(5,5;\bZ)  ~,
\ee
which has a single cusp at weak string coupling (the attractor mechanism in the presence in the background branes turns five of the moduli into fixed scalars~\rcite{Ferrara:1995ih,Seiberg:1999xz}).  The reduced U-duality group $\Gamma_{\bf q}\subset O(5,4;\bZ)$ implies that the elements $\gamma\in O(5,4;\bZ)$ that are not in $\Gamma_{\bf q}$ map this weak coupling cusp to another weak-coupling cusp of the moduli space~\rcite{Seiberg:1999xz,Larsen:1999uk}.  One can either regard this other cusp as a region of the moduli space with the same background charge and different moduli, or as having the same moduli and different charges.%
\footnote{An elementary example of this phenomenon is string theory on a circle, where string momentum and winding charges $(p,w)$ are a doublet under the $\bZ_2$ T-duality group.  In the presence of a winding string of winding $w=n_1$, the T-duality group is broken; the moduli space at fixed winding charge has two asymptotic regions, $R\to\infty$ and $R\to 0$.  Alternatively, one can divide the moduli space into two disjoint domains, with charges $(0,n_1)$ and $(n_1,0)$, both having the usual vacuum moduli space $R>1$.}
Adopting the latter interpretation, ${\bf q}'=\gamma\,{\bf q}$ is another charge vector having the same symplectic inner product $N$.
Thus each factorization $N=n_5n_1$ corresponds to a different weak-coupling cusp.  

The symmetric product orbifold lies in the cusp of the moduli space with charges $n_5=1,n_1=N$~\rcite{Larsen:1999uk}.  The RNS worldsheet formalism describes weakly coupled string theory in other cusps of the moduli space, having $n_5>1$.

\begin{figure}[ht]
\centering
\includegraphics[width=0.7\textwidth]{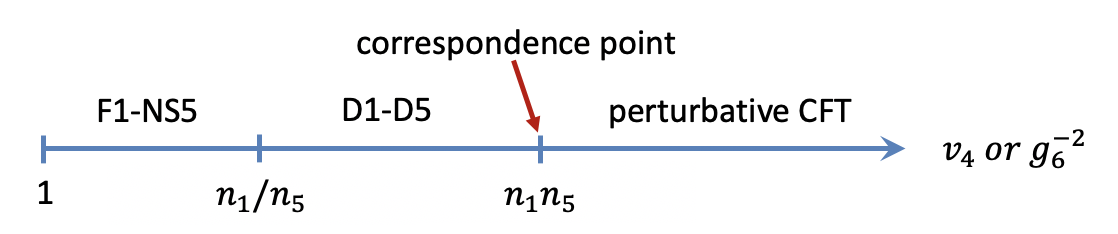}
\caption{\it Effective descriptions appropriate to various domains in the moduli space.}
\label{fig:modspace}
\end{figure}

Within a given cusp, the moduli space has several subdomains.  The low-energy string theory that applies is the one for which the fundamental string has the lightest tension among the branes that can wrap $\bT^4$.  In the NS5-F1 frame, the six-dimensional string coupling is one of the fixed scalars, pinned to
\be
g_6^2 = \frac{g_s^2 }{v_4} = \frac{n_5}{n_1}  ~,
\ee
while the compactification volume in string units $v_4$ is a modulus.  As one increases $v_4$, the string coupling increases until at $v_4=g_6^{-2}={\frac{n_1}{n_5}}$ ($g_s=1$), a D1-brane wrapping $\bT^4$ becomes as light as an F1 string wrapping the same cycle.  Beyond this point the S-dual D5-D1 frame is appropriate; in the process, the fixed scalar $g_6^{-2}$ and the modulus $v_4$ interchange roles.  Increasing the NS5-F1 frame volume $v_4$ further (\ie\ in the D5-D1 frame, increasing $g_6^{-2}$), one reaches a correspondence transition at $v_4 = n_5n_1$ beyond which the effective field theory on the D-branes becomes weakly coupled.  This sequence is depicted in figure~\ref{fig:modspace}.
%
%
We see that NS5-F1 tends to be the appropriate description over most of the supergravity regime for $n_5\ll n_1$, while the D5-D1 frame is appropriate over most of that regime when $n_5\sim n_1$.
Again, the symmetric product orbifold $\cM^N/S_N$, $\cM=\bT^4$ or $K3$, sits in the cusp with $n_5=1$, $n_1=N$, where the NS5-F1 description is appropriate (albeit stringy) in the entire range of validity of the low-energy bulk description.  Indeed, recently a worldsheet string theory realization of this regime has been proposed~\rcite{Eberhardt:2019ywk,Gaberdiel:2020ycd}.

\subsection{BPS ground states of the symmetric product orbifold}
\label{sec:Rgdstates}

The twisted sectors of the symmetric orbifold on a spatial circle $\bS^1_y$ are labelled by conjugacy classes of the symmetric group, corresponding to a choice of twisted boundary condition in which the $N$ copies of the SCFT on $\cM$ are partitioned into $N_\kappa$ groups of $\kappa$ copies each of $\cM$ which are cyclically sewn together, with $\sum_\kappa \kappa N_\kappa\tight=N$.  Each resulting $\kappa$-cycle is effectively a copy of the $\cM$ SCFT on a spatial circle $\kappa$ times longer.  In particular, the 1/2-BPS Ramond ground states are the same as those of a single copy of the SCFT; these are labeled by the cohomology of the target space $\cM$, with even (odd) cohomology associated to bosonic (fermionic) ground states.  One thus has eight bosonic and eight fermionic states for $\bT^4$ and 24 bosonic states for $K3$.  
Labeling the ground states of a single copy of $\cM$ by $I$, the Ramond ground states of the symmetric product are given by
\be
\label{symprodBPS}
\big| \Psi \big\rangle = \prod_{\kappa,I} \Big( \big| I \big\rangle_\kappa\Big)^{N_\kappa^I}
~~,~~~~
\sum_{\kappa,I} \kappa N_{k}^I = N ~.
\ee

Regardless of the effective supergravity description that applies in a particular domain, the 1/2-BPS spectrum is robust across the moduli space~-- the BPS states in any other regime of the moduli space can be written in a basis that uses the same labelling as that of the symmetric product.  The symmetric product describes a regime in which the appropriate effective description has a single fivebrane; then the cycle lengths $\kappa$ correspond to F1 winding $\kappa$.%
\footnote{This fact lies at the heart of the worldsheet description of $AdS_3\times\bS^3\times\cM$ at $n_5=1$ constructed in~\rcite{Eberhardt:2019ywk,Gaberdiel:2020ycd}.}
The cusps whose interpretation has $n_5>1$ can be labelled by the same data~\eqref{symprodBPS}.  However, in the regime where the number of fivebrane quanta in the effective supergravity is $n_5$, it is natural to interpret the cycle length $\kappa$ in~\eqref{symprodBPS} as a {\it fractional} string winding number, since F1 charge is only integral when $\kappa$ is a multiple of $n_5$~\rcite{Dijkgraaf:1996cv,Dijkgraaf:1997ku}.

Indeed, it has been proposed~\rcite{Dijkgraaf:1996cv,Maldacena:1996ya,Dijkgraaf:1997ku,Seiberg:1997zk} that the internal dynamics of a stack of $n_5$ NS5-branes is governed by a 6d self-dual string dynamics known as {\it little string theory}.  When a fundamental string is absorbed into the stack of fivebranes, it fractionates into $n_5$ constituent little strings.%
\footnote{The qualitative explanation of this phenomenon varies with the duality frame.  In type IIA, the little strings are realized in the M-theory lift of the strong-coupling region near the fivebranes as M2-branes stretching between M5-branes.  In type IIB, one can think of them as codimension four instantons in the effective 5+1d super Yang-Mills theory on the fivebranes.}
F1 winding $w$ then becomes little string winding $n_5 w$.

String winding fractionates momentum according to the winding quantum.  The 1/4-BPS excitations above these Ramond ground states consist of oscillator excitations on each $\kappa$-cycle, and the $\kappa$-cycle itself carries $\kappa/n_5$ units of fundamental string winding.  Thus the oscillator mode numbers (and the resulting $y$ momentum quantization) come in fractions of $\kappa$
\be
\label{symprod qtrBPS}
\big| \{n_\ell^{A\dot A} \},\{m_{j}^{B\beta} \},I \big\rangle_{\kappa\cdot\it  cycle} = 
\prod_{\ell,A\dot A;j,B\beta} (\alpha_{-\ell/\kappa}^{A\dot A})^{n_{\ell}^{A\dot A}} (\psi_{-j/\kappa}^{B\beta})^{m_{j}^{B\beta}} \big| I \big\rangle_\kappa ~.
\ee
Here $\alpha_p^{A\dot A}$ are modes of the $\bT^4$ currents, and $\psi^{B\beta}_p$ are their superpartners.  The full 1/4-BPS state of the symmetric product is then a symmtrized tensor product of such excited $\kappa$-cycles.

The entropy of these 1/4-BPS states will be discussed below in section~\ref{sec:stringyBPS}.


\section{Worldsheet setup} 
\label{sec:setup}

The gauged Wess-Zumino-Witten (WZW) model for the group quotient
\be
\label{coset}
\frac\cG\cH = \frac{\sltwo\times\sutwo\times \bR_t\times\bS^1_y}{U(1)_L\times U(1)_R} ~,
\ee
with $\cH$ consisting of a pair of null isometries of $\cG$, describes a family of BPS backgrounds of the NS5-F1 system~\rcite{Martinec:2017ztd,Martinec:2018nco,Martinec:2019wzw,Martinec:2020gkv,Bufalini:2021ndn}.  The target space of the worldsheet theory consists of this coset model times $\cM=\bT^4$ or $K3$ (the latter is most conveniently realized as the orbifold $\bT^4/\bZ_2$).

The radius $R_y$ of $\bS^1_y$ is a modulus of this background, which characterizes the crossover between a geometry which is approximately $[(AdS_3\times \bS^3)/\bZ_k] \times\cM$ at small radius, rolling over to a fivebrane throat $\bR_\rho\times\bR_t\times\bS^1_y\times\bS^3\times\cM$ with a linear dilaton in the radial coordinate~$\rho$ at large radius.  The discrete parameter $k$ characterizes the choice of embedding of $\cH\subset\cG$.  One can think of the crossover point $\rho\sim \half\log(\frac{kR_y}{n_5})$ as the charge radius of the background strings.  The full geometry is given in Appendix~\ref{sec:conventions}.  The $AdS_3$ decoupling limit is the limit $R_y\to\infty$.  The worldsheet theory describes perturbative excitations around a particular background 1/2-BPS state, whose bulk geometry is $(AdS_3\times \bS^3)/\bZ_k \times\cM$ in the $AdS_3$ decoupling limit.  The corresponding CFT state in the description~\eqref{symprodBPS} will be given below, after we develop more of the holographic map.

Physical vertex operators in worldsheet string theory on this background (for details of the construction, see~\rcite{Martinec:2020gkv}) lie in the cohomology of the BRST operator
\be
\label{BRSTop}
\cQ_{\rm\sst BRST} = \oint\!dz\, \big[ \big( cT + \gamma G +{\it ghosts} \big) + \big( \tilde c\cJ + \tilde\gamma \lambdab \big)\big]  
\ee 
and its right-moving counterpart.
The first set of terms implement the usual (super)reparametrization constraints, while second set implement the constraints under the gauged null currents $\cJ,\bar\cJ$ and their superpartners $\lambdab,\bar\lambdab$
\begin{align}
\begin{split}
\label{nullstuff}
\cJ = J^3_\sl + l_2 J^3_\su + l_3 \,i\partial t + l_4\, i\partial y
~~&,~~~~
\lambdab = \psi^3_\sl + l_2 \,\psi^3_\su + l_3 \,\psi^t + l_4\, \psi^y
\\[.2cm]
\bar\cJ = \bar J^3_\sl + r_2 \bar J^3_\su + r_3 \,i\bar\partial t + r_4\, i\bar\partial y
~~&,~~~~
\bar\lambdab = \bar\psi^3_\sl + r_2 \,\bar\psi^3_\su + r_3 \,\bar\psi^t + r_4\, \bar\psi^y  ~.
\end{split}
\end{align}
Our choice of null vector coefficients describing BPS supertubes is~\rcite{Martinec:2017ztd}
\be
\label{nullcoeffs}
l_2 = -1~,~~~ l_3=-l_4= - kR_y
\quad;\qquad
r_2 = -1~,~~~ r_3=r_4=-kR_y ~.
\ee
Our conventions on $\sltwo$ and $\sutwo$ current algebra and its representation theory, null gauging choices, \etc, largely parallel those of~\rcite{Martinec:2020gkv}, with some differences that are detailed in Appendix~\ref{sec:conventions}.  We set $\alpha'=1$.

The construction of vertex operators begins with a center of mass 
wavefunction 
\be
\label{comfn}
\Phi^{(w)}_{j;m,\mbar} \,
\Psi^{(w',\bar w')}_{j';m',\mbar'} \,
e^{ -iEt + iP_y y + i\bar P_{y} \bar y }
\ee
where $y(z),\bar y(\bar z)$ are the (anti-)holomorphic parts of the boson $y$; $\Phi^{(w)}_{j;m,\mbar}$ is an $\sltwo$ primary of the bosonic WZW model in the spectral flow sector $w$; and $\Psi^{(w',\bar w')}_{j';m',\mbar'}$ is a bosonic $\sutwo$ primary in the (L,R) spectral flow sector $(w',\bar w')$.  In particular $j,m,\mbar$ are quantum numbers under the bosonic $\sltwo$ currents $j^a_\sl$ and similarly $j',m',\bar m'$ under the bosonic currents $j^a_\su$.  The total currents \eg\ appearing in~\eqref{nullstuff} are then
\be
J_\sl^a = j_\sl^a - \frac i2 \T{(\epsilon_\sl)}{a}{bc}\psi_\sl^b \psi_\sl^c  
~~,~~~~
J_\su^a = j_\su^a - \frac i2 \T{(\epsilon_\su)}{a}{bc}\psi_\su^b \psi_\su^c  ~,
\ee
where the totally antisymmetric symbols have $\epsilon_\sl^{123}=\epsilon_\su^{123}=1$, and indices are raised and lowered with the relevant Killing metric.  We then denote the total spins by $J,J'$ respectively.
The (L,R) $y$-circle momenta are given by
\be
P_y = \frac{n_y}{R_y} + w_y R_y
~~,~~~~
\bar P_{ y} = \frac{n_y}{R_y} - w_y R_y  ~.
\ee
We specialize to vanishing $\bT^4$ momentum, as there are no BPS vertex operators carrying such momenta in the fivebrane decoupling limit~\rcite{Larsen:1999uk}.
Note that the $y$-circle momentum quantum $n_y$ is the contribution of the vertex operator to the conserved momentum charge $n_p$ carried by the system.

One then decorates these center-of-mass operators with oscillator excitations for the NS sector, or a spin field plus oscillator excitations for the R sector, and asks that they commute with the BRST operator.

We will largely work in the $(-1)$ picture for the $\beta\gamma$ ghosts in the NS sector, and the $(-\hf)$ picture for the R sector; we denote by $\varphi$ the scalar that bosonizes the ghost number current $\beta\gamma$.  There are analogous ghosts $\tilde\beta,\tilde\gamma,\tilde\varphi$ for gauging the fermionic null currents, which appear in the Ramond sector vertex operators (for the NS sector, one can work in the zero picture for these ghosts since there is no ghost number anomaly).


\subsection{The NS-NS sector} 
\label{sec:NSNS}

Supergravity vertex operators in the NS sector have a single fermionic excitation of each chirality.  There are 12 such fermions $\psi^a_\sl,\psi^a_\su,\psi_t,\psi_y,\psi^i_{\bT^4}$ for the left movers, which the BRST constraints winnow down to 8 physical polarizations; similarly for the right-movers.  These were analyzed in~\rcite{Martinec:2018nco,Martinec:2020gkv,Bufalini:2022wzu}. One can choose a gauge such that $w=0$, and we will do so in what follows.  The $\sutwo$ spectral flows parametrized by $w',\bar w'$ label inner automorphisms of the current algebra representations; nonzero values are realized as current algebra descendants (oscillator excitations), and so we can set $w'=\bar w'$ when discussing supergravity modes.

The four $\bT^4$ polarizations are manifestly transverse, leading to the left-moving vertex operator structure (here and below, we suppress the right-moving structure whenever possible to reduce notational clutter)
\be
\label{T4polns}
\cZ^{A\dot A}_{j,m;j',m'} = 
e^{-\varphi}\,\psi^{A\dot A}_{\bT^4}\,
\Phi^{(w)}_{j;m} \,
\Psi^{(w')}_{j';m'} \,
e^{ -iEt + i P_y y}
~~,~~~~ A,\dot A=\pm.
\ee
The mass-shell condition (the Virasoro zero mode constraint on $L_0\tight+\bar L_0$) sets $j=j'+1$.

The remaining four polarizations are most conveniently analyzed by projecting the products of $\sltwo$ and $\sutwo$ fermions with the c.o.m. wavefunction~\eqref{comfn} onto operators of fixed total spin 
\be
\label{NSshift}
J=j+\epsilon 
~~,~~~~
J'=j'+\epsilon' 
\quad,\qquad
\epsilon,\epsilon'=\pm1,0  ~,
\ee
and denote the resulting operators by
\be
\label{eps-shift-ns}
(\psi_\sl \Phi_j)_{j+\epsilon,m}
~~,~~~~
(\psi_\su \Psi_{j'} )_{j'+\epsilon',m'}
\ee
where $m,m'$ now refer to the total $J^3_\sl,J^3_\su$ quantum number rather than the bosonic one.
This is useful for the analysis of the BRST constraints because  the worldsheet supercurrent $G$ in the BRST operator~\eqref{BRSTop} is a singlet of the total spin.  
As a result, the zero mode of the left null constraint reads
\be
\label{leftnull-0}
0 \;=\; m+ l_2 m' +l_3 \frac E2 +l_4 \frac{P_y}2
\;=\; m-m' - \frac12 kR_y\Big( E- \frac{n_y}{R_y} - w_y R_y \Big)\;.
\ee

One thus trades the polarization labels $a,a'$ of $\psi^a_\sl,\psi^{a'}_\su$ for $\epsilon,\epsilon'$.  It turns out that $\epsilon=0$ and $\epsilon'=0$ lead to states which are either not BRST invariant, or are BRST exact.  One is left with the four physical polarizations~\rcite{Bufalini:2022wzu}
\begin{align}
\label{NSops}
\cW^{\epsilon}_{j,m;j',m'} &= 
e^{-\varphi}\Big[
\big(\psi_\sl\Phi_j\big)_{j+\eps,m}\,\Psi_{j',m'}
+\big(c^t_\eps\,\psi^t + c^y_\eps\psi^y\big) \Phi_{j,m}\,\Psi_{j',m'}
\Big] \, e^{-iEt+iP_y y}
\nn\\[.2cm]
\cX^{\epsilon'}_{j,m;j',m'} &= 
e^{-\varphi}\Big[
\Phi_{j,m}\,\big(\psi_\su\Psi_{j'}\big)_{j'+\eps',m'}
+\big(d^t_{\eps'}\,\psi^t + d^y_{\eps'}\psi^y\big) \Phi_{j,m}\,\Psi_{j',m'}
\Big] \, e^{-iEt+iP_y y}
\end{align}
labelled by the choices of $\eps,\eps'$.  The mass shell condition again sets $j=j'+1$.  The specific form of the Clebsches projecting onto definite spin in $\sltwo$ and $\sutwo$, as well as the values of $c^{t,y},d^{t,y}$, can be found in~\rcite{Bufalini:2022wzu}; in particular, one finds that the latter are of order $n_5/kR_y$ and so vanish in the $AdS_3$ limit $R_y\to\infty$.

Complete vertex operators combine one of the eight choices $\cV^i,\cW^\eps,\cX^{\eps'}$ for left-movers with an independent choice $\bar\cV^i,\bar\cW^{\bar\eps},\cX^{\bar\eps'}$ for right-movers.


\subsection{The R-R sector} 
\label{sec:RR}

The left-moving part of a supergravity R-R vertex operator takes the form
\be
\cY^{\vareps_1...\vareps_6} = 
e^{-\half\varphi+\half\tilde\varphi} \,
S^\perp_{\vareps_1\vareps_2\vareps_3} \,
S^{||}_{\vareps_4\vareps_5\vareps_6} \,
\Phi_{j,m-\half\vareps_1}\Psi_{j',m'-\half\vareps_2} \,
e^{-iEt+iP_y y}  ~,
\ee
where
\be
S^\perp_{\vareps_1\vareps_2\vareps_3} = 
e^{\frac i2(\vareps_1H_1+\vareps_2H_2+\vareps_3H_3)}
~~,~~~~
S^{||}_{\vareps_4\vareps_5\vareps_6} =
e^{\frac i2(\vareps_4H_4+\vareps_5H_5+\vareps_6H_6)}
\ee
are spin fields for $AdS_3\times \bS^3$ and $\bR_t\times\bS^1_y\times\bT^4$, respectively. 
Our bosonization conventions set 
\begin{align}
\label{eq:bosonizations}
\begin{split}
\psi^\pm_\sl &= e^{\pm iH_1}
~,~~~
\psi^\pm_\su = e^{\pm iH_2}
~,~~~
\psi^3_\su\pm\psi^3_\sl = e^{\pm iH_3}
\\[.2cm]
\psi^6\pm i\psi^7 &= e^{\pm iH_4}
~,~~~
\psi^8\pm i\psi^9 = e^{\pm iH_5}
~,~~~
\psi^y\pm\psi^t = e^{\pm iH_6} ~,
\end{split}
\end{align}
where directions 6,7,8,9 span $\bT^4$.
The mass-shell condition again sets $j=j'+1$.  We choose a GSO projection 
\be
\label{eq:GSO-L}
\prod_{\alpha=1}^6 \vareps_\alpha = -1 
\ee
in order that the 10d GSO projection on physical states turns out to select positive chirality 10d spinors in the $AdS_3$ decoupling limit $R_y\to\infty$.
We will also find it useful to define the $AdS_3\times\bS^3$ chirality
\be
\vareps = \vareps_1\vareps_2\vareps_3
\ee
and eliminate $\vareps_3,\vareps_6$ via
\be
\label{fixeps}
\vareps_3=\vareps\vareps_1\vareps_2
~~,~~~~
\vareps_6=-\vareps\vareps_4\vareps_5 ~.
\ee
One can characterize operators by their leading terms in the large $R_y$ ($AdS_3$) limit, in which   one again trades the polarizations $\vareps_1,\vareps_2$ for projections onto definite $\sltwo$ and $\sutwo$ spins 
\be
\label{spinshift}
J=j+\eps ~~,~~~~
J'=j'+\eps' \quad,\qquad 
\eps,\eps'=\pm\half~.
\ee
The physical RR vertex operators in the $AdS_3$ limit can then be written as
\begin{align}
\label{Rops}
\cY^{\vareps,\vareps_4;\eps,\eps'}_{j,m;j',m'} = 
e^{-\half\varphi+\half\tilde\varphi} \,
\big(S^\perp\Phi_j\Psi_{j'}\big)^\vareps_{j+\eps,m;j'+\eps',m'} \, S^{||}_{\vareps_4,\vareps_5=\vareps\vareps_4} \,
e^{-iEt+iP_y y} + O(1/R_y)
\end{align}
with $\vareps_6=-1$ determined by~\eqref{fixeps} via the solutions to the fermionic null constraint.  For details, see~\rcite{Bufalini:2022wzu}.%
\footnote{Note that our conventions here for null vector coefficients $l_2,r_2,l_4,r_4$ have the opposite signs compared to the choice made in~\rcite{Bufalini:2022wzu}.  This results in a flip in the signs of $\vareps_3,\vareps_6$ in the solutions to the constraints.}

The fermionic constraints from Virasoro and null gauging leave eight physical left-moving polarizations.  One finds that, for the leading terms~\eqref{Rops} in $1/R_y$, these have $\vareps=+$ for $\eps=-\eps'$ (and $\vareps=-$ for $\eps=+\eps'$) for any choice of $\vareps_4$.  These comprise the eight left-moving physical Ramond polarizations; one has a similar set for the right-movers.

We should note that the flipped sign $r_4=-l_4$ for the right-movers leads to a flipped sign for $\bar\vareps_6$ in the solution to the constraints.  This leads to the opposite choice for GSO projection in 12d:
\be
\label{eq:GSO-R}
\prod_{i=1}^6 \bar\vareps_i = +1 ~.
\ee


\section{1/2-BPS spectrum} 
\label{sec:hfBPS}

The backgrounds~\eqref{coset} with the gauged null currents specified in~\eqref{nullcoeffs} describe particular 1/2-BPS Ramond ground states in the spacetime CFT.  We now consider vertex operators that preserve the same supersymmetries as the background.  These spacetime supersymmetries are described in Appendix~\ref{sec:BPScondition}.  Maximally BPS vertex operators carry winding that contributes to the F1 charge, but no $\bS^1_y$ momentum as this would break another half of the supersymmetry.

The vertex operators among~\eqref{NSops}, \eqref{Rops} that preserve the background supersymmetry have
\be
n_y=0 ~~,~~~~ E = w_y R_y  ~.
\ee 
The left null constraint $\cJ=0$, equation\;\eqref{leftnull-0}, 
then imposes~\rcite{Martinec:2018nco,Martinec:2020gkv}
\be
\label{leftnull}
0 = 
m - m' = -\sfn -\epsilon - \sfm + \epsilon' - 1 ~,
\ee
where we define $\sfm$ and $\sfn$ through%
\footnote{The choice of conjugate discrete series representations $\cD_j^-$ for $\sltwo$ corresponds to string creation operators.}
\be
\label{j3vals}
m=-J-\sfn  ~~,~~~  \sfn=0,1,2,\dots  %
\quad;\qquad
m' = -J' + \sfm ~,~~~ \sfm = 0,1,\dots,2J'+1~,
\ee
and recall $j=j'+1$.
The only solutions are $\sfm=\sfn=0$ with $\epsilon'=\epsilon+1$, giving rise to the choices
\be
\label{Clebsches}
\epsilon=-1,~\epsilon'=0
\quad;\qquad
\epsilon=0,~\epsilon'=+1
\quad;\qquad
\epsilon=-\epsilon'=-\half ~.
\ee

The BPS polarization states are then
\begin{align}
\label{halfBPS}
\cV^+_{j',w_y} &\equiv \cW^-_{j'+1,-j';j',-j'}
\nn\\[.2cm]
\cV^-_{j',w_y} &\equiv \cX^+_{j'+1,-j'-1;j',-j'-1}
\\[.2cm]
\cS^A_{j',w_y} &\equiv \cY^{\vareps=+,\vareps_4=\vareps_5=A;\eps=-,\eps'=+}_{j'+\half,j'+\half;j'+\half,-j'-\half}  
~~,~~~~  A=\pm~. \nn
\end{align}
There are thus two NS and two R left-BPS polarizations, for each value of $\sutwo$ c.o.m. spin $j'$.  The Clebsches of the polarization vector with the center-of-mass wavefunction result in the total spins $J=J'$
\begin{align}
\label{BPSspins}
\cV^+ ~&:~~ J = j-1 = j' ~, ~~ J'=j'
\nn\\[.2cm]
\cV^- ~&:~~ J = j = j'+1 ~, ~~ J'=j'+1
\\[.2cm]
\cS^A ~&:~~ J = j-\half = j'+\half ~, ~~ J'=j'+\half ~.
\nn
\end{align}

1/2-BPS operators are of the form~\eqref{halfBPS} on both left and right.  Among the bosonic operators, there are four NS-NS and four R-R operators.  As we recall in the next subsection, these match the deformations of supertubes, the 1/2-BPS geometries of the NS5-F1 system.  This deformation spectrum is easy to understand via T-duality along $\bS^1_y$, which converts the background into NS5-P.%
\footnote{Note that this T-duality is trivial in the worldsheet theory, corresponding to a flip in the relative sign between $l_4$ and $r_4$ in the null gauge currents~\rcite{Martinec:2017ztd}.}
The excitations are now BPS momentum waves on the fivebrane~-- four transverse scalars $X^{\alpha\dot\alpha}$
(in a bispinor labeling of the transverse $\bR^4$),
whose mode excitations are related to 
\be
\label{Vaadot}
\cV^{\alpha\dot\alpha}_{j',w_y}=\cV^\alpha_{j',w_y}\bar\cV^{\dot\alpha}_{j',w_y}~,
\ee
and four polarizations of the type IIA NS5 gauge multiplet (a scalar and a self-dual antisymmetric tensor) whose mode excitations are related to
\be
\label{SAB}
\cS^{AB}_{j',w_y} = \cS^A_{j',w_y}\bar\cS^B_{j',w_y} ~.
\ee  
The internal gauge excitations are R-R because they carry the flux sourced by D-branes which can end on NS5-branes.  After the T-duality, $w_y$ is now the momentum quantum on the T-dual circle, and the vertex operator describes a supergravity mode in spacetime, but this is simply a relabelling of the worldsheet data.

For $K3$ realized as $\bT^4/\bZ_2$, there are an additional 16 R-R gauge modes coming from the orbifold fixed point cohomology~\rcite{Harvey:1995rn}, and sourced by D-branes wrapping the orbifold vanishing cycles~\rcite{Douglas:1996sw,Diaconescu:1997br}.


\subsection{Nonlinear deformation: Supertubes} 
\label{sec:supertubes}

The 1/2-BPS supergravity solutions sourced by these excited fivebranes were enumerated in~\rcite{Lunin:2001fv,Lunin:2001jy,Taylor:2005db,Kanitscheider:2007wq}.  
The metric is given by
\be
\label{LMbkgd}
d s_{10}^2 = 
-\frac{Z_5}{\cP}\Bigl[\big(du\!\!\,+\!\omega \big) \big(dv\!\!\,+\!\beta\big) \Bigr] 
+ Z_5 \,ds_{\!\perp}^2 +  \,ds_{\cM}^2   ~,  
\ee
where
\begin{equation}
\cP   \equiv     Z_1 \, Z_5  -  Z_0^2 - Z_{(\gamma)}^2 ~.
\label{curlyP}
\end{equation}
Here $ds^2_{10}$ is the ten-dimensional string-frame metric; $ds_\perp^2$ is the metric on the space $\bR^4$ transverse to the branes, parametrized by $\xx^{\alpha\dot\alpha}$; $ds^2_\cM$ is the metric on the $\bT^4$ or $K3$ compactification; and we denote $u,v=t\tight\pm y$.
For the rest of the supergravity fields, see Appendix~\ref{sec:STgeom}; and for further discussion, see~\rcite{Skenderis:2006ah,Kanitscheider:2006zf,Giusto:2013rxa,Bena:2015bea,Giusto:2015dfa,Giusto:2019qig}.

The harmonic functions and forms appearing in the geometry are expressed in terms of a set of source functions $\sfF^{\alpha\dot\alpha}(\hat v), \sfF^{AB}(\hat v)$ for the bosonic supergravity fields (and $\sfF^{\alpha B}(\hat v),\sfF^{A\dot\alpha}(\hat v)$ for fermions), through a set of Green's function integrals
\begin{align}
Z_5 &= \frac{n_5}{L}\int_0^{L} \frac{d\hat v}{|{\bf x}^{\alpha\dot\alpha}-{\mathsf F}^{\alpha\dot\alpha}(\hat v)|^2} \hfill
\nn\\[.2cm]
{\mathsf A}_{\alpha\dot\alpha} &= \frac{n_5}{L}\int_0^{L} \frac{d\hat v \,\dot \sfF_{\!{\alpha\dot\alpha}}(\hat v)}{|{\bf x}-{\mathsf F}(\hat v)|^2}
~~,~~~~
d{\mathsf B} = *_{\!\scriptscriptstyle \perp} d{\mathsf A}
~~,~~~~
\beta=\sfA+\sfB~,~~\omega = \sfA-\sfB
\label{LMints}\\[.2cm]
Z_1 &= 1 + 
\frac{n_5}{L}\int_0^{L} \frac{d\hat v \, (\dot{\mathsf F}^{\alpha\dot\alpha}  \dot{\mathsf F}_{\hskip -1pt {\alpha\dot\alpha}}+\dot\sfF^{AB}\dot\sfF_{\!AB})}{|{\bf x}-{\mathsf F}(\hat v)|^2}
~~,~~~~
\nn\\[.2cm]
Z_{(I)} &= \frac{n_5}{L}\int_0^{L} \frac{d\hat v \,\dot \sfF_{\!(I)}(\hat v)}{|{\bf x}-{\mathsf F}(\hat v)|^2}  
~~,~~~~  I = 0,\gamma~; ~~~~ \gamma= 1,...,b_2^-  ~,
\nn
\end{align}
where $b_2^-$ is the rank of the anti-selfdual middle chomology of $\cM$.
For $\bT^4$, we can relabel the $Z_{(I)}$ into a bispinor $Z_{AB}$, where $0=[AB]$ is the antisymmetric singlet and $\gamma=1,2,3$ comprise the symmetric triplet $(AB)$.  In what follows, we mostly concentrate on this case, and mention differences for $K3$ where appropriate.

The polarizations $\sfF^{\alpha\dot\alpha}$ specify the location of the fivebranes in their transverse space.  To bind all the fivebranes together, one imposes a twisted boundary condition so that the fivebrane charge is realized by a single fivebrane that wraps the $y$-circle $n_5$ times.  

\begin{figure}[ht]
\centering
  \begin{subfigure}[b]{0.4\textwidth}
  \hskip 0cm
    \includegraphics[width=\textwidth]{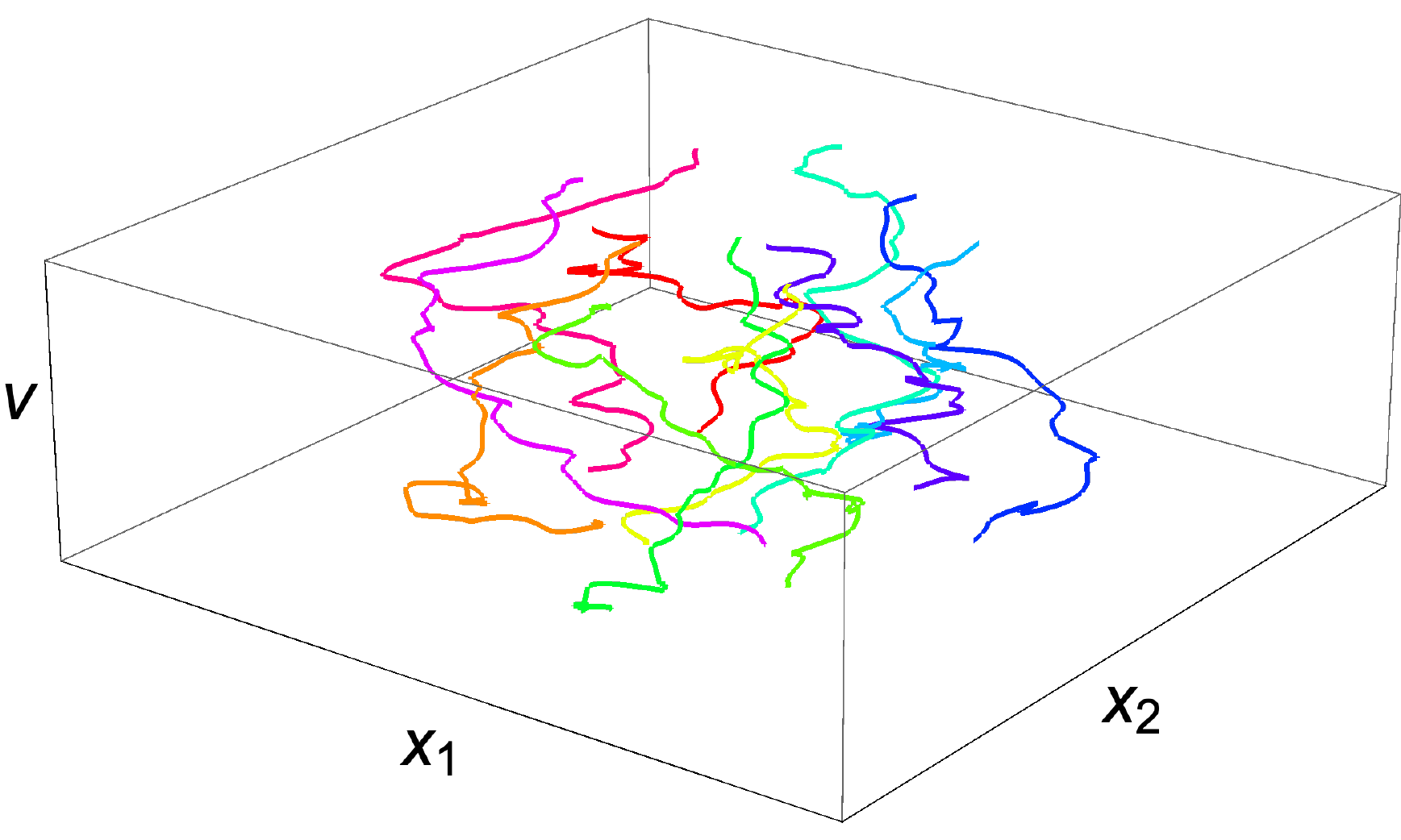}
    \caption{ }
    \label{fig:ThermalPrimary}
  \end{subfigure}
\qquad\qquad
  \begin{subfigure}[b]{0.4\textwidth}
      \hskip 0cm
    \includegraphics[width=\textwidth]{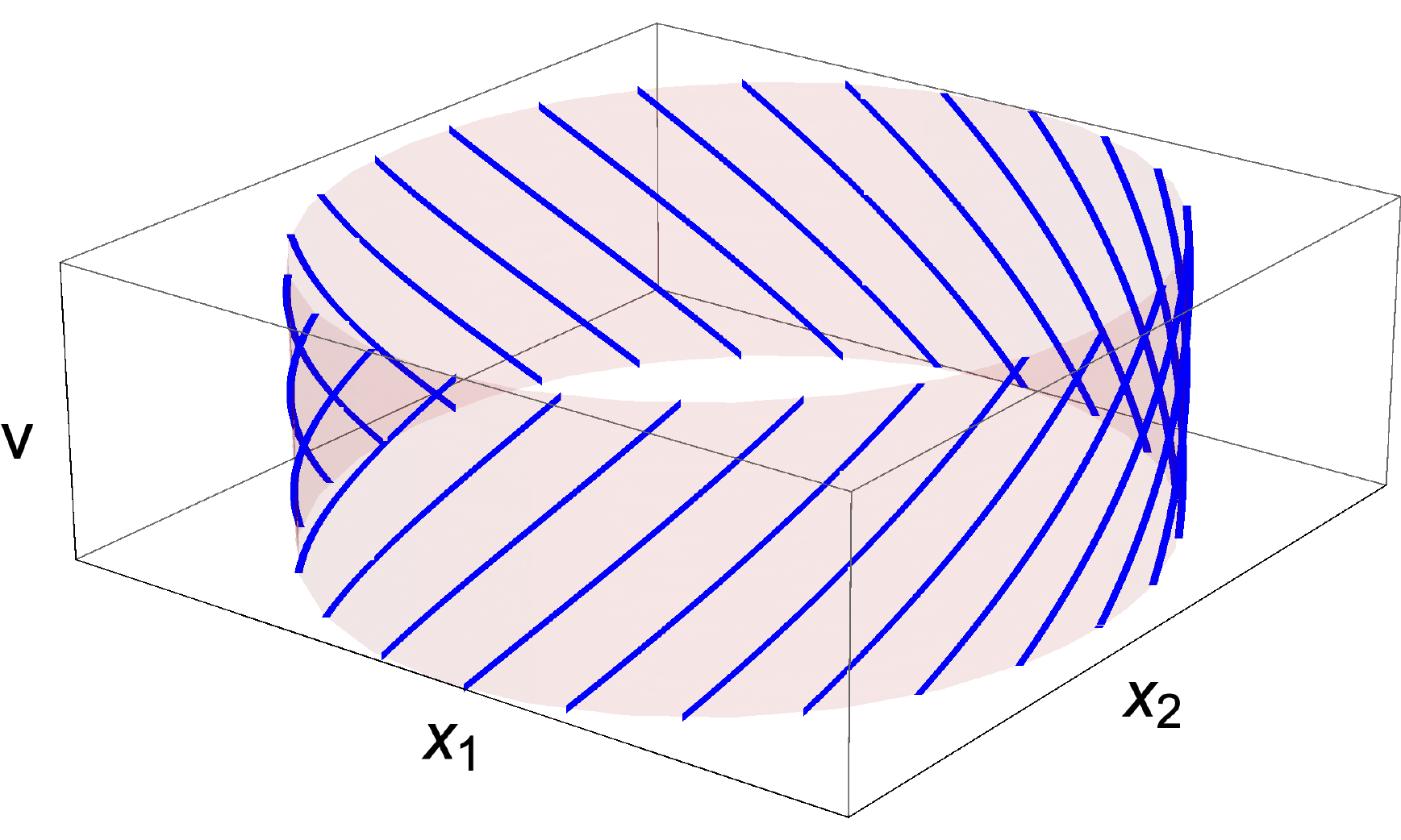}
    \caption{ }
    \label{fig:CircularST}
  \end{subfigure}
\caption{ 
(a) {\it Typical source profile $\sfF({\hat v})$, with successive windings along $\bS^1_y$ color coded with evolving hue around the color wheel to indicate their connectivity.}
(b) {\it Circular supertube source profile, in which only a single mode is excited (in this case, $k=3$ and $n_5=25$), so that the fivebranes spiral around a torus in $(y,x^1,x^2)$ shaded in pink.}
}
\label{fig:Ellipses}
\end{figure}

This structure is depicted in figure~\ref{fig:ThermalPrimary}.  This figure should be thought of as describing the T-dual NS5-P source configuration, where the momentum waves on the fivebrane indeed specify the wiggling shape of the fivebrane in its transverse space.  One should then mentally T-dualize this picture to NS5-F1 (in particular we have $L=2\pi n_5/R_y$).  Since the fivebrane source is partly along and partly transverse to the circle being dualized, the dual geometry has a local KK monopole structure (there is no net KKM charge).  The separation of the fivebranes along the dual $\tilde y$-circle is coded in $B$-fluxes through two-cycles in the KKM structure.  For instance, in figure~\ref{fig:CircularST} there are $k=3$ NS5's vertically along the $y$-circle, leading to three coincident KKM's after the T-duality; this is one way to see why the NS5-F1 geometry sourced by the profile is $(AdS_3\times\bS^3)/\bZ_k$ (with the fivebrane separation along the dual $\tilde y$-circle transforming into $B$-fluxes through the vanishing cycles, rendering the orbifold non-singular in string theory).

A basis of states is specified by the set of occupation numbers $\{N^{\alpha\dot\alpha}_p,N^{AB}_p \}$ for the Fourier modes of these source functions (and $\{N^{\alpha B}_p,N^{A\dot\alpha}_p \}$ for their fermionic superpartners)
\be
\label{groundstate}
\big|\Psi\big\rangle = \prod_{\substack{p,q,r,\ell\\ \alpha,\dot\alpha,A,B}}
\Bigl(\,|\alpha\dot\alpha\rangle\strut_{p}\Bigr)^{N^{\alpha\dot\alpha}_p}
\Bigl(\,|AB\rangle\strut_{q}\Bigr)^{N^{AB}_q}
\Bigl(\,|\alpha B\rangle\strut_{r}\Bigr)^{N^{\alpha B}_r}
\Bigl(\,|A \dot\alpha\rangle\strut_{\ell}\Bigr)^{N^{A \dot\alpha}_\ell}
~~.
\ee
For the extension to fermionic modes, see~\rcite{Taylor:2005db}. 
The eight bosonic and eight fermionic polarization states match those of the symmetric product~\eqref{symprodBPS}.

In the T-dual NS5-P frame, the $N^I_p$ are simply the mode occupation numbers for a BPS wave on a single fivebrane wrapping the $y$-circle $n_5$ times, with modes labelled by $p$ contributing a fractional amount $p/n_5$ to the quantized $y$-momentum due to the multiple covering.  In the NS5-F1 frame, the corresponding F1 modes carried on the fivebrane have fractional string winding $p/n_5$, and so might be interpreted as modes of the fractionated little string that is thought to underlie the dynamics of decoupled fivebranes.

The worldsheet formalism transpires in the grand canonical ensemble of fixed chemical potential for F1 winding, rather than fixed winding~\rcite{Porrati:2015eha}.  The Legendre transform to fixed winding imposes the constraint 
\be
N \equiv n_1n_5 = p N_p^{\alpha\dot\alpha}+q N_q^{AB} + r N_r^{\alpha B} + \ell N_\ell^{A\dot\alpha} ~,
\ee
but we will work in the ensemble that is natural to the worldsheet.
A vertex operator with winding $w_y$ on the $y$-circle introduces/extracts that winding at the spacetime boundary.  On the other hand, the vertex operators can also redistribute the string winding already in the initial state among the various mode numbers.

The round supertube background described by the null-gauged WZW model is a coherent state built on a single mode%
\footnote{The worldsheet formalism works in the grand canonical ensemble with respect to F1 winding\rcite{Kim:2015gak}.}
\be
\label{roundST}
\big| \Psi_{\textrm bkgd} \big\rangle = \sum_{N_k^{++}} \frac{(a_k)^{N_k^{++}}}{(N_k^{++})!} \Big(\big|\alpha\dot\alpha\tight=+\!+\!\big\rangle_k \Big)^{N_k^{++}} ~,
\ee
so that $\langle N \rangle = k\langle N_k^{++} \rangle = ka_k$.  
This source profile is depicted in figure~\ref{fig:CircularST}.
In~\rcite{Martinec:2020gkv}, it was argued that the 1/2-BPS operators~\eqref{halfBPS} extract $2j'\tight+1$ background modes, in addition to adding F1 winding $w_y$ (which due to mode fractionation adds $w_yn_5$ to the mode number), resulting in a single mode with winding
\be
\label{fracwind}
(2j'\tight+1)k+w_y n_5  ~.
\ee
At the same time, they implement a change in the polarization state according to~\rcite{Martinec:2020gkv}
\begin{align}
\begin{split}
\label{transition}
\cV^{\alpha\dot\alpha}_{j',w_y}~&:~~
\big(| \!+\!+\rangle_k \big)^{2j'+1}
\longrightarrow~
\big| \alpha\dot\alpha\big\rangle_{(2j'+1)k+w_yn_5}
\\[.2cm]
\cS^{AB}_{j',w_y}~&:~~
\big(| \!+\!+\rangle_k \big)^{2j'+1}
\longrightarrow~
\big| AB\big\rangle_{(2j'+1)k+w_yn_5}  ~.
\end{split}
\end{align}
The identification of these vertex operators as mediating these transitions is consistent with the conservation of string winding charge on $\bS^1_y$, as well as the difference in $J^3_\su,\bar J^3_\su$ charges between the LHS and RHS.

Note that, in order for the background to consist of a single (multiply wound) fivebrane source, $k$ and $n_5$ must be relatively prime (otherwise, one has $\gcd(k,n_5)$ interleaved fivebrane sources).  Therefore, by adjusting $j',w_y$ one can generate any desired mode number of fractional F1 winding from applying these operators to the background, in the regime where perturbative string theory applies.  More precisely, one can generate any mode number that is not a multiple of $n_5$; the range of $j'$ is 
\be
\label{jprange}
j'= 0,1,\dots, \hf n_5-1
\ee
and so such mode numbers cannot be realized in perturbative string theory.  Little string winding that is a multiple of $n_5$ corresponds to integer F1 winding; such modes lie at the threshold of a continuum of long string states, and as such their wavefunctions are not normalizable~\rcite{Argurio:2000tb}.%
\footnote{In other words, these fivebrane modes mix with those of unbound F1 string states.  These modes are in normalizable bound states at generic moduli, but are unbound at the codimension four locus in moduli space described by the worldsheet formalism, which has all the R-R moduli set to zero.  The fractionally moded strings, on the other hand, are always bound to the fivebranes.}

The exponentiation of these vertex operators into the worldsheet action formally generates the complete ensemble of 1/2-BPS backgrounds over a finite domain of their configuration space.  This is particularly clear when the R-R fields are turned off; purely NS backgrounds 
yield worldsheet nonlinear sigma models that are conformal to all orders in sigma mode perturbation theory (\ie\ to all orders in the $\alpha'$ expansion)~\rcite{Horowitz:1994rf,Tseytlin:1996yb} (see in particular the discussion around equations 7-9 of the second reference).

Effects such as the near-source geometry and effects such as the location of the fivebranes on the T-dual to the $y$-circle, are captured by non-perturbative properties of the sigma model.  In the subset of configurations where the background is purely NS and sourced by fivebrane wiggling in a two-dimensional plane (of which the background~\eqref{roundST} is an example), these non-perturbative aspects are captured by a dual superpotential, in a generalization of the Calabi-Yau/Landau-Ginsburg correspondence~\rcite{Ooguri:1995wj,Giveon:1999px,Giveon:1999tq,Giveon:2016dxe,Martinec:2020gkv}.  We will review this construction in section~\ref{sec:singularities}.


\section{1/4-BPS supergravity spectrum} 
\label{sec:superstrata}

Among the 1/4-BPS deformations of the round supertube, there are a large number of supergravity vertex operators.  These arise from combining (say) any of the four BPS polarizations~\eqref{halfBPS} among the right-movers, with an arbitrary polarization state~\eqref{NSops}, \eqref{Rops} among the left-movers.  In particular, the center-of-mass contribution to the vertex operator need no longer be highest weight in $\sltwo$ or $\sutwo$.  
The vertex operators~\eqref{NSops}, \eqref{Rops} pick a particular Clebsch~\eqref{NSshift} or~\eqref{spinshift}, which need not be the same for the left- and right-movers.
The BPS constraint on the right forces
\be
\bar m = -\bar J = -(j + \bar\epsilon)
~~,~~~~
\bar m'= -\bar J' = -(j' + \bar\epsilon') ~.
\ee
and recall the mass shell condition sets $j=j'+1$.
The axial ($L_0-\bar L_0$) Virasoro constraint is the usual level-matching requirement
\be
\label{axialVir}
n_y w_y + m' w' - \bar m'\bar w' +\frac{n_5}4\big[(w')^2-(\bar w')^2\big] + N_L-N_R =0
\ee
where $N_{L,R}$ are the left- and right-moving oscillator excitation levels, and we work in the zero spectral flow sector $w=0$ for $\sltwo$.  For a supergravity operator, there is no $\sutwo$ winding $w'=\bar w'=0$, and $N_L=N_R=\half$ in the $(-1)$ picture NS sector; similarly in the R sector one has a ground state spin field on both sides in the $(-\half)$ picture.  
The constraint then requires either $n_y=0$ or $w_y=0$.
The axial ($\cJ-\bar\cJ$) null gauge constraint then imposes
\be
\label{axialnull}
\sfm+\sfn + (\epsilon-\bar\epsilon) - (\epsilon'-\bar\epsilon') = k n_y  ~,
\ee
which will require $n_y>0$ if $\sfm+\sfn\ne 0$ (a case-by-case analysis below shows that the net effect of the $\epsilon$'s on the LHS is a non-negative contribution).
Thus, in order to turn on a 1/4-BPS supergravity deformation by exciting the center-of-mass zero modes on the left, we must have no winding on the $y$-circle,
\be
w_y = 0 ~.
\ee

Interestingly, for the 1/2-BPS vertex operators, one must set $n_y=0$ but is allowed $w_y\ne 0$, while for 1/4-BPS supergravity vertex operators one has $w_y=0$ but nonzero $n_y$.  The difference lies in the roles played by these two sets of operators.  On the one hand, the 1/2-BPS operators deform the winding string condensate in the background, and thus in general carries winding on the $y$-circle (special cases, where the winding is a low multiple of $k$, may be generated solely by extracting strings from the background, but for general winding one needs nonzero $w_y$).  On the other hand, the 1/4-BPS deformations add a supergravity wave on top of the winding string condensate, carrying momentum on the $y$-circle; in order to be a supergravity excitation rather than an excited string state, the operator cannot have both winding {\it and} momentum along $\bS^1_y$.


\subsection{Nonlinear deformation: Superstrata} 
\label{sec:qtrBPS}

Just as the exponentiation of 1/2-BPS vertex operators into the action coherently deforms the background to a nearby supertube background, the exponentiation of the 1/4-BPS supergravity vertex operators deforms the background into a nearby {\it superstratum} background.  These are smooth supergravity solutions carrying all three charges NS5-F1-P; for instance, the metric~\eqref{10dmetric} generalizes to
\be
ds^2 = -\frac{2Z_5}{\cal P}\Bigl[(dv\!\!\,+\!\beta)\big(du\!\!\,+\!\omega - \hf \cF (dv\!+\!\beta)\big) \Bigr] 
+ Z_5 \,ds_{\!\perp}^2 +  \,ds_{{\tiny\mathbb T}^4}^2
~~,~~~~
\cP = Z_1Z_5-Z_{AB}^2 ~,
\ee
with similar generalizations for the other fields~\rcite{Giusto:2013rxa,Bena:2017xbt}.

The most studied family of superstratum solutions consider supergravity backgrounds that excite 6d tensor supermultiplets in addition to the 6d gravity supermultiplet.  Originally, solutions involving the same four NS and four R polarizations as one has for the 1/2-BPS deformations~\eqref{Vaadot}, \eqref{SAB}
were considered~\rcite{Bena:2015bea,Bena:2016ypk,Bena:2017xbt}.  Let us denote these operators by
\be
\label{BPSpolns}
\cV^{\alpha\dot\alpha}_{j';\sfn,\sfm}
~~,~~~~
\cS^{AB}_{j';\sfn,\sfm}  ~,
\ee
where $\sfm,\sfn$ denote the c.o.m. excitations discussed above.

Because the polarization state remains the same, the vertex operators implement the same transitions~\eqref{transition}, but there are now left-moving excitations above the ground state in the final state.
The corresponding excitations of the symmetric product were identified in~\rcite{Bena:2015bea,Bena:2016ypk,Bena:2017xbt} as cycles of the form%
\footnote{More precisely, initially only solutions built on excitations of the ground state $\eps^{AB}|AB\rangle\equiv|00\rangle$ were considered, as these preserve more symmetry and are more straightforward to construct explicitly.  More recently, the generalization to other ground states has been analyzed in~\rcite{Ganchev:2021iwy}.}
\begin{align}
\label{origstrata}
\big| \sfm,\sfn;I\big\rangle_{(2j'+1)k} = \big(\sfJ_{-1/k}^+ \big)^\sfm \big( \sfL_{-1/k}- \sfJ^3_{-1/k} \big)^{\sfn}  \big| I\big\rangle_{(2j'+1) k}
\end{align}
where the polarization state is $I=\alpha\dot\alpha$ for the NS-NS sector and $I=AB$ for the R-R sector; and $\sfL,\sfJ$ are modes of the spacetime superconformal algebra.%
\footnote{For $k>1$, a special case 
\be
\sfn=0~~,~~~~\sfm=pk~~,~~~~2j'\tight+1=pk~~,~~~~ n_y=p
\nn
\ee
of these deformations was considered in~\rcite{Bena:2016agb}.  The general case was discussed in~\rcite{Shigemori:2020yuo}.}
The vertex operators~\eqref{BPSpolns} again implement transformations that conserve $y$-circle winding as well as $J^3_\su,\bar J^3_\su$.  The excitation structure matches as well, in that the $\sfn$ units of excitation of the zero mode vertex operators~\eqref{BPSpolns} are implemented by the global $\sltwo$ lowering operator $(J^-_\sl)_0$, which corresponds to $\sfL_{-1}$ in spacetime; similarly the $\sfm$ units of excitation in $\sutwo$ are implemented by $(J^+_\su)_0$, which maps to $\sfJ^+_{-1}$ in the Ramond sector of the spacetime CFT.

It was subsequently realized that in order to solve the BPS equations in the presence of multiple modes of the form~\eqref{BPSpolns}, one must expand the set of deformations to include their bosonic superpartners (so-called ``supercharged'' modes~\rcite{Ceplak:2018pws,Heidmann:2019zws,Rawash:2021pik}) having the analogous polarization structure, but choosing the other Clebsch in $\sltwo$ and $\sutwo$ (\ie\ switching the signs of $\eps,\eps'$ in~\eqref{NSops}, \eqref{Rops}).  Each of the two supercharges applied to the highest weight of the supermultiplet acts to raise the $\sltwo$ spin and lower the $\sutwo$ spin.  The resulting left-moving polarization states (lying in the same supermultiplet as the corresponding modes~\eqref{BPSpolns}) thus have the total spins
\begin{align}
\label{supchged}
\widehat\cV^- \equiv \cW^+ ~&:~~ J = j+1 = j'+2 ~, ~~ J'=j'
\nn\\[.2cm]
\widehat \cV^+ \equiv \cX^- ~&:~~ J = j = j'+1 ~, ~~ J'=j'-1
\\[.2cm]
\widehat \cS^A \equiv \cY^{\vareps=+,\vareps_4=A;\eps=+,\eps'=-} ~&:~~ J = j+\half = j'+\frac32 ~, ~~ J'=j'-\half
\nn
\end{align} 
leading to another set of vertex operators
\be
\label{supchgd}
\widehat\cV^{\alpha\dot\alpha}_{j';\sfn,\sfm} = \widehat\cV^{\alpha}_{j';\sfm,\sfn} \bar\cV^{\dot\alpha}_{j',w_y=0}
~~,~~~~
\widehat\cS^{AB}_{j';\sfn,\sfm} =
\widehat\cS^A_{j';\sfn,\sfm} \bar\cS^{B}_{j',w_y=0}  ~.
\ee
Because the contribution of $\epsilon-\epsilon'$ to the axial null constraint~\eqref{axialnull} changes sign from $-1$ to $+1$, this constraint now reads
\be
\label{supch axnull}
\sfm + \sfn + 2 = k n_y ~,
\ee
implying that the corresponding symmetric product CFT state has two additional units of left-moving momentum excitation relative to~\eqref{origstrata}.  Similarly, in the R-R vertex operator the Clebsches of the spin field $S$ with the center-of-mass operator $\Phi_j\Psi_{j'}$ has the opposite sign in both $\sltwo$ and $\sutwo$ relative to~\eqref{halfBPS}, once again leading to~\eqref{supch axnull}.

The symmetric product description of the supercharged modes is given by~\rcite{Ceplak:2018pws,Heidmann:2019zws,Rawash:2021pik}
\begin{align}
\label{supchg}
\Big(\sfG^{+1}_{-\frac1k} \sfG^{+2}_{-\frac1k}+\frac1k \sfJ^+_{-\frac1k} \big(\sfL_{-\frac1k}-\sfJ^3_{-\frac1k}\big)\Big)\, \big|\sfm,\sfn;I\big\rangle_{(2j'+1)k}
\end{align}
where $\big|\sfm,\sfn;I\big\rangle_{(2j'+1)k}$ is given in~\eqref{origstrata}, and $\sfG$ is the spacetime CFT supercurrent.
Indeed, the two additional (fractional) units of left-moving momentum match the quantum numbers of the vertex operator~\eqref{supch axnull}. 

Note also that we have flipped the assignments between $\widehat\cV^\pm$ and $\cW^+,\cX^-$ in~\eqref{halfBPS}. \eqref{BPSpolns} relative to $\cV^\pm$ and $\cW^-,\cX^+$ in~\eqref{supchgd}.  This choice is motivated by the difference of $\sutwo$ spins between the initial and final states~-- the value of $J'$ for $\cW^+$ is one less than $\cX^+$, and similarly that of $\cX^-$ is one less than $\cW^-$.  This corresponds to the fact that the mode~\eqref{supchg} has $\sutwo$ spin which is one more than that of~\eqref{origstrata} due to the application of the two $\sfG$'s.

All told, these deformations comprise 8 NS-NS and 8 R-R deformations for each choice of $j',\sfm,\sfn$ allowed by~\eqref{j3vals},~\eqref{jprange} (half each from the original superstratum modes, and half each from the supercharged modes).  These comprise half of the 32 bosonic 1/4-BPS supergravity modes.  The remainder are associated to 6d vector multiplets, as we now describe.


\subsection{Nonlinear deformation: 6d vector multiplets} 
\label{sec:vectors}

The remaining 1/4-BPS supergravity deformations are modes of 6d vector multiplets.  Half of these are straightforward to describe~-- they are NS-NS operators that combine a BPS polarization $\bar\cV^\pm$ of~\eqref{halfBPS} for the right-movers with one of the $\bT^4$ polarizations $\cZ^{A\dot A}$ of~\eqref{T4polns} for the left-movers.  In addition, there are eight more R-R vector modes that arise when the 6d chiralities of the spin fields are opposite on left and right, $\bar\vareps=+$ for the BPS right-movers and $\vareps=-$ for the non-BPS left-movers.%
\footnote{Note that the excitations charged under the NS-NS vectors are momentum and F1 winding on $\bT^4$.  For the R-R vectors, the left and right spin fields have opposite $\bT^4$ chirality and so comprise odd rank antisymmetric tensors that couple to D1 and D3-branes entirely wrapped on $\bT^4$.  For type IIA, the $\bT^4$ chiralities are flipped: $\bar\vareps=+,\vareps=-$ correspond to the same $\bT^4$ chirality, leading to even rank antisymmetric tensors coupling to D0, D2 and D4-branes wrapping the torus.}

The NS-NS vectors have $\sltwo$ and $\sutwo$ total spins
\begin{align}
\begin{split}
\label{NSvec}
\cZ^{A\dot A} \bar \cV^+ ~&:~~
-m=j\tight+\sfn=j'\tight+1\tight+\sfn ~,~~~ m'=-j'\tight+\sfm ~,~~~ -\bar m=j-1=j' ~,~~~ \bar m'=-j'
\\[.2cm]
\cZ^{A\dot A} \bar \cV^- ~&:~~
-m=j\tight+\sfn=j'\tight+1\tight+\sfn ~,~~~ m'=-j'\tight+\sfm ~,~~~ -\bar m=j=j'+1 ~,~~~ \bar m'=-j'-1
\end{split}
\end{align}
These modes belong to the NS-NS sector and are clearly 6d vectors~-- the worldsheet NS-NS supergravity vertex operators represent perturbations of $(G+B)_{LR}$; the above modes have the R index in 6d and the L index on the $\bT^4$ (or $K3$) compactification, and so dimensionally reduce to 6d vectors. 

The axial null constraint sets
\be
\label{vecnull}
\sfm+\sfn+1 = kn_y ~. 
\ee
These quantum numbers match those of the symmetric product cycle where a ground state is excited by a single supercurrent applied to a fermionic ground state
\be
\label{NSvec symprod}
\big| \sfm,\sfn;\dot\alpha,A\dot A \big\rangle_{(2j'+1)k} = 
\big(\sfJ_{-1/k}^+ \big)^\sfm \big( \sfL_{-1/k}- \sfJ^3_{-1/k} \big)^{\sfn}
\sfG_{-1/k}^{\alpha\dot A } \big|A\dot\alpha\big\rangle_{(2j'+1)k}
~~,~~~~
\alpha=+
\ee
which carries the quantum numbers $A\dot A$ of a vector on $\bT^4$ coming from the left-moving sector. 
Note that the $\sutwo$ spin of~\eqref{NSvec symprod} is the same as for the NS modes~\eqref{origstrata} for $\alpha=+$.  This is consistent with the fact that~\eqref{NSvec} and $\cV^{+\dot\alpha}_{j';\sfn,\sfm}$ have the same $\sutwo$ spin.  In other words, the F1 winding and $\sutwo$ spins are consistent with the proposed identification of the transition of 1/4-BPS states implemented by the vertex operators~\eqref{NSvec}
\be
\cZ^{A\dot A}\bar\cV^{\dot\alpha}~:~~~~ 
\Big(\big|\!+\!+\rangle \Big)^{2j'+1}
~\longrightarrow~ 
\big| \sfm,\sfn;\dot\alpha,A\dot A \big\rangle_{(2j'+1)k}  ~.
\ee

In the R-R sector, the remaining physical vertex operators that are right-BPS consist of
\begin{align}
\label{RRvec}
\cY^{\vareps=-,\vareps_4=\dot A;\eps=\eps'}
\bar \cS^A ~:~~
-m&=j\tight+\frac\eps2\tight+\sfn=j'\tight+1\tight+\frac\eps2\tight+\sfn 
~,~~~ m'=-j'\tight-\frac\eps2\tight+\sfm ~,~~~ 
\\[.2cm]
-\bar m&=j\tight-\half=j'\tight+\half ~,~~~ \bar m'=-j'\tight-\half ~.
\nn
\end{align}
The natural candidate for the corresponding symmetric product state is again a supercurrent acting on a fermionic ground state
\be
\label{RRvec symprod}
\big| \sfm,\sfn;\beta,B\dot A \big\rangle_{(2j'+1)k} = 
\big(\sfJ_{-1/k}^+ \big)^\sfm \big( \sfL_{-1/k}- \sfJ^3_{-1/k} \big)^{\sfn}
\sfG_{-1/k}^{\alpha\dot A} \big|\beta B \big\rangle_{(2j'+1)k}
~~,~~~~  \alpha=+
\ee
(with $\beta=\epsilon$) which again carries a vector index on the $\bT^4$, but now the bispinor $\dot AB$ comes half from the left-movers and half from the right-movers as one expects for a R-R operator.%
\footnote{Note that the left- and right-moving NS/R parities of a vertex operator correspond to left and right fermion parities in the spacetime CFT.  Furthermore, we assign fermion parity $(-1)^F\tight=+1$ to the transverse ground state polarizations $\alpha,\dot\alpha$ and $(-1)^F\tight=-1$ for the internal polarizations $A,B$ (on $\bT^4$, the fermion zero modes implement transitions between these ground states and thus fix these assignments).  Thus~\eqref{NSvec symprod} has even fermion parity on both left and right, while~\eqref{RRvec symprod} has odd fermion parity on both sides.} 
The $\alpha,\beta$ $\sutwo$ spins combine to make $J_L-J_R$ equal to zero or one as in~\eqref{RRvec}.  Again one can check that all the remaining conserved quantum numbers are compatible with the identification of the vertex operator~\eqref{RRvec} as mediating the transition from $\big(\ket{++}\big)^{2j'+1}$ to~\eqref{RRvec symprod}. 

The $\sutwo$ spins of the symmetric product states~\eqref{RRvec symprod} are either one more than ($\beta=+$), or the same as ($\beta=-$) that of the R-R states in~\eqref{origstrata}.  In other words, we identify the transition implemented by~\eqref{RRvec} as having the final state~\eqref{RRvec symprod}, with $\epsilon=\beta$.

These R-R fields are also 6d vectors.  In type IIB, the R-R fields are even rank antisymmetric tensors in 10d; the vertex operators~\eqref{RRvec} are vectors on the internal space, and thus odd rank antisymmetric tensors in 6d~-- either vectors or three form potentials that are electric-magnetic duals of vector potentials.

The 1/4-BPS 6d vector perturbations~\eqref{NSvec}, \eqref{RRvec} can be exponentiated into smooth, nonlinear deformations of the background~\rcite{Ceplak:2022wri}.  With only gravitational and 6d tensor multiplet perturbations, one can cast the BPS field equations as a three-step hierarchy of linear equations, with the unknown harmonic forms in each subsequent layer of the hierarchy having sources bilinear in harmonic forms solved for in previous layers.  The core idea is that the coefficients of homogeneous solutions to the lower level equations can be adjusted to ensure the smoothness of the solutions to higher level equations.  The work of~\rcite{Ceplak:2022wri} extends this structure to include 6d vector multiplets as well as tensor multiplets; the hierarchy of linear equations now has five layers, but otherwise the structure is similar.  

For compactification on $K3$, there are no 6d vectors in the effective supergravity theory.  In the symmetric orbifold, there are no fermionic ground states on which to build the states~\eqref{NSvec symprod} or~\eqref{RRvec symprod}.  On the worldsheet, if one realizes $K3$ as $\bT^4/\bZ_2$, the vertex operators~\eqref{NSvec}, ~\eqref{RRvec} are projected out by the $\bZ_2$ orbifold.  Instead, one has an additional 16 tensor multiplets in 6d~-- one for each of the supersymmetric ground states arising from the fixed points of $\bT^4/\bZ_2$ and their orbifold cohomology.  These will lead to an additional set of 1/4-BPS RR vertex operators coming from the use of the BPS polarizations on both left and right, and their supercharged counterparts where one flips the Clebsches on the left.


\section{Stringy 1/4-BPS spectrum} 
\label{sec:stringyBPS}

The general 1/4-BPS vertex operator allows both $n_y$ and $w_y$ to be nonzero, and combines a right-moving vertex operator
\be
\bar\cV^{\dot\alpha}_{j',w_y,n_y}
~~,~~~~
\bar\cS^{B}_{j',w_y,n_y}
\ee
generalizing~\eqref{halfBPS} with a general left-moving vertex operator
\be
\label{generalBPS}
e^{-\varphi} \,
{\cal P}\big(\textit{L~osc.}\big)\, \Phi_{j+1,m}^{sl} \,\Psi_{j',m',w'}^{su} 
\,e^{-iEt+iP_y y}
\ee
for the NS sector, and similarly for the Ramond sector, subject to the Virasoro and null gauging BRST constraints.  Here $\cP$ is a polynomial in (derivatives of) the currents and their superpartners.  The axial Virasoro constraint~\eqref{axialVir} determines the level $N_L$ of the left-moving oscillator excitations.

One sees that the momentum carried by the oscillator excitations is fractionated by $1/w_y$, as opposed to the c.o.m. excitations, which are fractionated by $1/k$ due to the axial null constraint~\eqref{axialnull}, \eqref{vecnull}.  Note that neither of these fractionations reaches the maximum one might expect based on the symmetric product cycles~\eqref{symprod qtrBPS} for the corresponding cycle length~\eqref{fracwind}, by a factor of order $n_5$ for the longest cycles (noting the range of $j'$, equation~\eqref{jprange}).  

This shortfall in momentum fractionation is perhaps not so surprising, as these 1/4-BPS operators are adding fundamental strings to the background rather than fractional ``little string'' excitations. The surprise is that the 1/2-BPS vertex operators~\eqref{Vaadot}, \eqref{SAB} were able to capture generic fractional winding.  This was possible due to the $n_5$-fold winding of the background profile $\sfF^{\alpha\dot\alpha}(v)$ in~\eqref{LMints}.  The worldsheet physical state constraints do not allow a simultaneous fractionation of both momentum and winding beyond what one would expect from a fundamental string on a $\bZ_k$ orbifold spacetime, and/or wrapping a circle.

It could be that some states with the most finely fractionated excitations lift off the BPS bound as one moves from the symmetric orbifold point to the supergravity regime, but it is unlikely that all of them do since they are essential to explaining black hole entropy.  It may just be that states with such highly fractionated excitations are not realized among the microstate geometries and perturbatively stringy microstates.

It is natural to conjecture that a coherent excitation of 1/4-BPS strings~\eqref{generalBPS} would lead to something along the lines of a geometry with an explicit F1-P macroscopic string source as in~\rcite{Dabholkar:1990yf}, but now in an ambient spacetime with $AdS_3\times\bS^3$ asymptotics.  One imagines that a suitable generalization of the superstratum construction might exist, where one relaxes the condition of complete smoothness of the geometry in favor of allowing perturbative string singularities sourcing momentum and winding charge.

Note that the operators~\eqref{generalBPS} do not create index states.  The supergravity 1/4-BPS states enumerated in the previous section saturate the elliptic genus, up to level $n_1n_5/4$~\rcite{deBoer:1998us}.  Thus none of these stringy BPS states are protected as we move around the CFT moduli space, at least up to this level.  Nevertheless, they lie on the BPS bound at least at this point in the moduli space and at tree level in string perturbation theory.  In this respect, they join the majority of 1/4-BPS supergravity states, which also vastly outnumber index states (comparing~\eqref{light-EG} to~\eqref{light-ent}; see for instance~\rcite{Benjamin:2016pil,Shigemori:2019orj}).

The growth of the elliptic genus at low level $1\ll n_p\ll 6N$ in the $K3$ symmetric product is given by~\eqref{light-EG}
(at $J_L=N/2$ in the R sector, equivalently $J_L=0$ in the NS sector).
At levels up to $N/4$, this is also the growth of the elliptic genus in supergravity.
A more general quantity known as the ``Hodge elliptic genus''~\rcite{Kachru:2016igs,Benjamin:2016pil}
\be
Z_{\rm\sst HEG} = \tr\Big[(-1)^{F_L+F_R}q^{L_0-\frac{c}{24}} y^{J_0} u^{\bar J_0} \Big]
\ee
(the elliptic genus is $Z_{\rm HEG}(u=1)$) is not in general an index, but counts BPS states; in supergravity, it has (at the same point in $J_L$) the growth~\rcite{Benjamin:2016pil} 
\be
\rho^{\rm\sst HEG}_{\rm sugra}(n_p) \sim 
\begin{cases}
N\, \exp\Big[\frac{4\pi}3 \big(12\,n_p^3\big)^{1/4} \Big]
&K3\\[.3cm]
N\, \exp\Big[\frac{4\pi}3 \big(8\,n_p^3\big)^{1/4} \Big]
&\bT^4
\end{cases}
\ee
while in the symmetric product it has Hagedorn growth~\rcite{Benjamin:2016pil}
\be
\label{symprod-HEG}
\rho^{\rm\sst HEG}_{\rm S^N(\cM)}(n_p) \sim \exp\big( 2\pi n_p \big)  ~.
\ee
An ansatz explaining this growth in the symmetric product BPS density of states was given in~\rcite{Bena:2011zw} (and generalized to include the effects of angular momentum).  Suppose that the cycles of the symmetric product are split into two groups~-- a set of $\ell$ short cycles of winding $k$ that carry the angular momentum, and a long cycle that carries the more entropic excitations.  Subtracting the portion of the charges residing in the short strings, the long string entropy is
\be
S_{\rm long} = 2\pi\sqrt{(n_1n_5-\ell k)n_p - (J_L-\ell/2)^2} ~.
\ee
Extremizing with respect to $\ell$, one finds $\ell=2(J_L-k n_p)$, and so
\be
\label{symprod enigma}
S_{\rm long} = 2\pi\sqrt{n_p(n_p k^2 +n_1n_5-2J_L k)} ~,
\ee
which recovers the result~\eqref{symprod-HEG} at $J_L=n_1n_5/2$ for $k=1$, and generalizes it to general angular momentum and short string length.  The fact that this result differs from the enigmatic black hole entropy~\eqref{enigma-ent} shows that these states indeed do move onto and off of the BPS bound as we move around the moduli space of the theory.

One also sees an analogous Hagedorn growth from the {\it perturbative} 1/4-BPS string spectrum.  The Fock space of perturbative strings has a symmetric product structure, with the restriction that the winding comes in the form $(2j'\tight+1)k+w_y n_5$ (with $0\le j'\le \half n_5-1$).  Thus the BPS strings don't have windings that are a multiple of $n_5$, as mentioned above around~\eqref{jprange}; for large $n_5$, this is a relatively minor restriction.  The oscillator spectrum is gapped by $1/w_y$ rather than the cycle length in symmetric product terms, which is approximately $n_5$ times longer at large $w_y$.  The worldsheet $L_0\tight-\bar L_0$ constraint~\eqref{axialVir} shows that the angular momentum subtracts from the available oscillator energy in much the same way that it does for the long cycles in the symmetric product (here we have as usual set $j=j'+1$).
One more effect~-- there are twice as many oscillator polarizations on a perturbative string as compared to the little string.  Assuming that the entropy is carried by a single long perturbative string, one has a winding budget $n_1n_5 = \ell k+ (2j'\tight+1)k+w_y n_5$ and an angular momentum budget $J_L = \half \ell + m'+\half n_5 w'$ (noting that the $\ell$ background cycles have length $k$ in little string units, and angular momentum~$1/2$).  Again extremizing with respect to $\ell$, and for simplicity ignoring the small effect of $j'$ for large $n_y,w_y$, one finds a perturbative 1/4-BPS string entropy
\be
\label{pert enigma}
S_{\rm pert} = 2\pi\sqrt{\frac{2}{n_5}n_p\big(n_p k^2 +n_1n_5-2J_L k\big)} ~.
\ee
Partitioning the entropy into several long strings leads to a similar result.

Not surprisingly, the extra factor of $n_5$ in the fractionation of little string excitations relative to those of fundamental strings contributes to the larger BPS entropy~\eqref{symprod enigma} relative to~\eqref{pert enigma} for $k=1$.   For $k>1$, the energy of the string redshifts by a factor $k$; and so perturbatively around the orbifold point $J_L=\frac{n_1n_5}{2k}$, the entropy is $S=2\pi\sqrt{2/n_5} \, E_{\rm local}$, where $E_{\rm local}=kn_p$ is the local energy around the orbifold cap of the geometry.

Of course, this comparison is not really appropriate, since the two are evaluated at different points in the moduli space of the theory.  The proper comparison is between~\eqref{pert enigma} and the dominant black hole ensemble~\eqref{enigma-ent}.  
Comparison of these two entropies at $J_L=\frac{n_1n_5}{2k}$ shows that (in the regime of validity $2J_L>n_p$) the enigma entropy dominates; for larger $n_p$ the BTZ entropy~\eqref{BHent} dominates.  Thus there is no regime where the density of states is dominated by perturbative strings, and thus no correspondence transition in the spectrum of 1/4-BPS states in this system.


\section{Singularities and non-singularities} 
\label{sec:singularities}


\subsection{Nonlinear (non)instabilities} 
\label{sec:stability}

An analysis performed in~\rcite{Eperon:2016cdd} suggested that the microstate geometries~\eqref{LMbkgd}, \eqref{LMints} might be classically unstable towards developing a singularity, through a nonlinear process whereby the system traps excitations near the ``evanescent ergosurface'', the supertube locus where the deepest redshift occurs.  Further investigation using an analysis of 1/2-BPS shockwave deformations of the supertube~\rcite{Marolf:2016nwu} showed that the suggested non-linear instability is nothing more than evolution in the supertube configurations space, and argued that this evolution would proceed until the supertube reached a generic configuration. 

The considerations here and in~\rcite{Martinec:2020gkv} allow us to see what is going on in stringy detail.  The analysis of~\rcite{Eperon:2016cdd} considered a limit of large $\bS^3$ angular momentum $j'$.  In this limit, the center-of-mass wavefunction of string vertex operators
\be
\label{localized}
\Phi_{j;-j,-j}^{\sl} \,\Psi_{j';-j',-j'}^{\su}
= e^{2ij\tau+2ij'\phi}\,\Big(\frac{a^2}{r^2+a^2}\Big)^{j}\, \sin^{2j'}\theta
\ee
localizes on the the fivebrane source at $r=0,\theta=\frac\pi2$.%
\footnote{This result generalizes to the class of 3-charge geometries also analyzed in~\rcite{Eperon:2016cdd}; see Appendix~\ref{sec:GLMTshocks}.}
As we saw above, the effect of the vertex operator is to lower the $\sutwo$ spin of the system by $j'$, by changing the moding~\eqref{roundST} according to~\eqref{transition}.
The $\sutwo$ spin $j'$ is bounded by $\half n_5-1$ according to~\eqref{jprange}; the mass shell condition sets $j=j'+1$.

In the supergravity limit where $n_5$ is macroscopic, one can consider macroscopic $j'$ and the perturbation can be extremely well localized near the ``evanescent ergosurface''.  This is the situation analyzed in~\rcite{Eperon:2016cdd}.  At the same time, the perturbation looks like a shockwave in this limit.  We can write the perturbed source profile as
\begin{align}
\begin{split}
\label{deformed}
\sfF^{\alpha\dot\alpha}(v) &= a_k \, \delta^{\alpha\dot\alpha,++}\, e^{ikv/n_5} + \sum_{j'}  f^{\alpha\dot\alpha}_{j'} \, e^{i(2j'+1)kv/n_5}
\\[.2cm]
\sfF^{AB}(v) &= \sum_{j'} f^{AB}_{j'} \, e^{i(2j'+1)kv/n_5} 
\end{split}
\end{align}
R-R deformations $f^{AB}$ are somewhat simpler to consider, since they don't change the location of the fivebrane source.  The contributions to the harmonic functions $Z_{AB}$ are evaluated in Appendix~\ref{sec:LMints}; they are highly oscillating along the source, as well as being highly localized there according to~\eqref{localized}.  In addition, there is a contribution to the onebrane harmonic function $Z_1$ via the stress tensor of the perturbation in the numerator of the integrand in~\eqref{LMints}.  If the product of perturbations has no low-frequency components other than the zero mode, this deformation is also highly localized near the supertube source, and yields a slight change in the radius of the supertube from the zero mode, plus a high-frequency perturbation.  If we ignore the high-frequency terms, we find precisely the shockwave solutions of~\rcite{Chakrabarty:2021sff}. 

For NS-NS perturbations, the transverse position of the fivebrane is modified by high-frequency wiggles, so the Lunin-Mathur integrals are more complicated.  Nevertheless, one finds much the same result~-- the solution is only modified near the source, apart from the contribution of the stress tensor zero mode to the winding charge.  One can see this via coarse-graining over the high-frequency wiggles in the source introduced by the deformation.

Thus, we see that the limit of large angular momentum analyzed in~\rcite{Eperon:2016cdd} is a variant of the shockwave limit analyzed in~\rcite{Marolf:2016nwu,Chakrabarty:2021sff}.  Both are part of a larger story about how bulk 1/2-BPS perturbations modify the state of the system, and how the correspondence between the bulk string theory and the CFT is manifested in the worldsheet vertex operators.  The phenomenon observed in~\rcite{Eperon:2016cdd} is not an instability so much as an indication that it costs no energy for the system to evolve along the 1/2-BPS configuration space.  The perturbations they analyze are simply a special case of~\eqref{transition}.  
The evolution described in~\rcite{Marolf:2016nwu} has the system shedding angular momentum; the entropy increases as the angular momentum decreases, and the larger phase space of final states drives the evolution in this direction.  Note that to shed angular momentum, it must be radiated away, which costs energy.  There is thus no instability of this sort for the decoupled system.  However, if we excite the isolated system away from the BPS bound, it will evolve along the configuration space at fixed angular momentum, at a rate governed by the available energy.

We also see that the shockwave singularity is resolved by stringy effects.  The worldsheet dynamics places constraints on the localizability of excitations.  We don't expect to be able to make string states localized to less than the string scale.  In the $AdS_3$ context, the $AdS$ curvature radius in units of the string length is the level of the $\sltwo$ WZW model
\be
k_\sl = \Big(\frac{R_{AdS}}{\lstr}\Big)^2
\ee
Normalizable discrete series affine representations $\cD^\pm_{j}$ have $\sltwo$ spins bounded by
\be
\half < j < \frac{k_\sl+1}2 ~.
\ee
Since the corresponding zero-mode wavefunctions behave as in~\eqref{localized}, the wavefunction can't be localized to a region smaller than of order the string scale, and to achieve that resolution involves $j\sim O(k_\sl)$.

There is furthermore a stringy property of affine $\sltwo$ representation theory (see~\rcite{Giveon:2016dxe,Martinec:2020gkv} for recent discussions and further references) that results in an identification of representations in adjacent spectral flow sectors
\be
\label{dualreps}
\cD^{\pm}_{j,w} \equiv \cD^{\mp}_{\half k_\sl+1-j, w\mp1}  ~,
\ee
where $w$ is a ``winding'' (spectral flow) quantum number.  What this means is that the wavefunction for a given pointlike string state with winding zero, having a wavefunction scaling as $r^{-2j}$ at large radius, has another branch of the wavefunction involving strings that wind once around the $AdS_3$ azimuthal direction, scaling at large radius as $r^{-(\half k_\sl+1-j)}$.  The two scalings exchange dominance at $j\sim \frac14 k_\sl$; for larger spins, the winding component of the wavefunction is more delocalized and at $j=\half(k_\sl+1)$ the wound string merges with a continuum of radially unbound strings in plane-wave states having $j=\half+is$.

Thus, the localizability of string states bounces between the $AdS$ scale for $j$ near a zero or $\half k_\sl$, and the string scale for $j\sim \frac14 k_\sl$.  If we want a localized shock, we should take $j$ large but smaller than of order $\frac14 k_\sl$ and take the large $k_\sl$ limit.  Since in the critical dimension we have $k_\sl=n_5+2$ for the level of the bosonic $\sltwo$ WZW model, we have to take the limit of a large number of fivebranes.

Naively, one might have hoped to increasingly localize the source by devoting an ever larger fraction of the winding budget to deformations $f^I$ in~\eqref{deformed}, noting the constraint~\eqref{Q1int}, \eqref{Q1Q5-n1n5}.  However, as we increase the winding on the $y$-circle, or equivalantly in $\sltwo$, the localizability of the shock is always bounded by the string scale.  Stringy effects (non-perturbative in $\alpha'$ but leading order in $g_s$) resolve the shockwave singularity.

To summarize, ground state deformations localize to the extent possible in the most redshifted parts of the geometry, where they backreact to affect the source configuration by changing the string condensate carried by the fivebranes.  The long-term trapping of these deformations seen in supergravity is simply a manifestation of motion along the 1/2-BPS configuration space.  The shockwave limit is one where the deformation has no low-frequency components; stringy effects resolve the shockwave over distances of order the string scale.

While this evolution takes the system along generically nonsingular configurations, singularities can arise at particular points in the configuration space.  We discuss these next.


\subsection{Singularities at fivebrane intersections} 
\label{sec:intersections}

\begin{figure}[ht]
\centering
\includegraphics[width=1.0\textwidth]{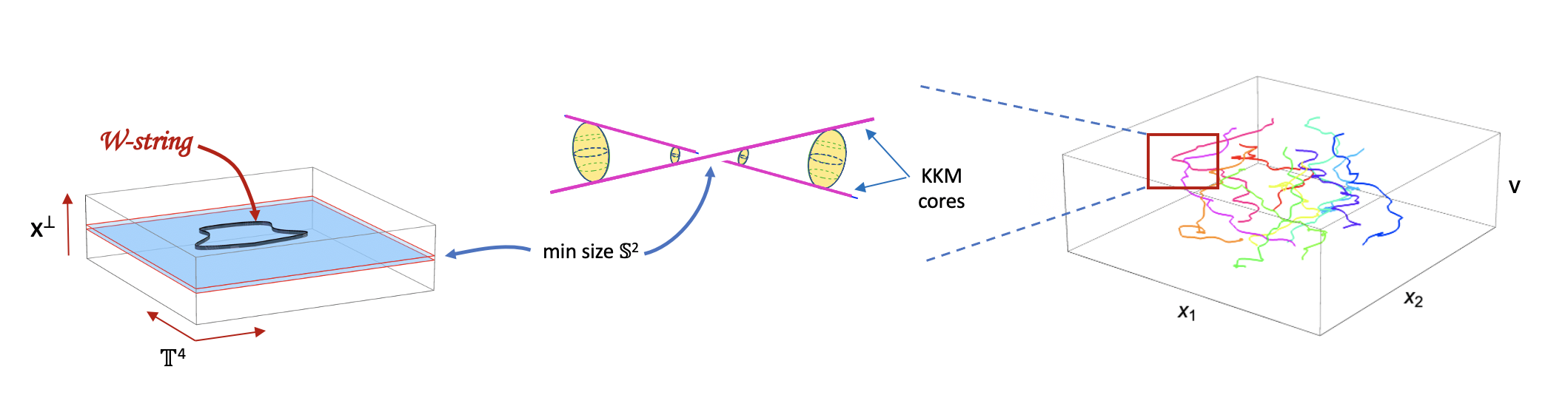}
\caption{\it Zooming in on fivebrane singularities.  {\rm Right figure:} The wiggly fivebrane source profile may come close to self-intersecting; the profile specifies the location of a codimension four KK monopole core where a fibered circle degenerates.  {\rm Middle figure:} The fibered circle together with the interval between strands of the source contour form the azimuthal and polar directions of a local $\bS^2$ which has minimal area at the near-intersection.  {\rm Left figure:} D3-branes wrapping this $\bS^2$ have their remaining leg wiggling along the $\bT^4$ compactification, making an effective tensionless string when the profiles do self-intersect~-- the W-string of little string theory. }
\label{fig:intersection}
\end{figure}

In section~\ref{sec:supertubes}, we saw that the geometry has a collection of KK monopole cores extending along a one-dimensional contour (these are the T-duals of NS5 windings in the NS5-P frame).  A pair of windings along the contour forms a two-sphere in which the KKM fibered circle forms the azimuthal direction, and the interval between the two windings forms the polar direction.  The size of this two-sphere is governed by the separation of the two strands of the source profile, and vanishes when the strands intersect.  See figure~\ref{fig:intersection}.  
Recall that the figure depicts source profile in the T-dual NS5-P frame, with the separation of the strands along $\bS^1_{\tilde y}$ in the figure indicating the amount of NS $B$-flux through this two-sphere in the NS5-F1 frame supertube.  

%
%

Let us exhibit this structure in a bit more detail.
Isolating the vicinity of a nearly self-intersecting profile as in the middle figure of figure~\ref{fig:intersection}, one can locally approximate each component of the source as a linear density $\kappa_i$, $i=1,2$ of fivebranes along a line in the transverse $\bR^4$.  Let us consider the first of these two line sources.  Let $x_1$ parametrize the line, which sits at the origin in the transverse $\bR^3$ parametrized by $x_{2,3,4}$; we work in spherical coordinates $(\rho,\vartheta,\varphi)$ in this $\bR^3$.  The harmonic functions arising from this line source are
\be
\label{harmfns}
Z_5 = \frac{\kappa}{\rho}
~~,~~~~
Z_1 = \frac{1}{\kappa\rho}
~~,~~~~ 
\sfA = \frac{dx_1}{\rho}
~~,~~~~
\sfB = \cos\vartheta \, d\varphi
\ee
so that the metric is
\be
ds^2 = \kappa\Big[ -\rho \, dt^2 - 2 dt\, dx + \rho \big( dy + \cos\vartheta\, d\varphi \big)^2 + \frac{d\rho^2}{\rho} +\rho \,d\Omega_2^{\,2} \Big] + ds^2_{\bT^4}  ~.
\ee
The geometry along the directions $(y,\rho,\vartheta,\varphi)$ is that of a Kaluza-Klein monopole, with $y$ parametrizing the fibered circle that shrinks away at $\rho=0$.  There is a similar structure for the other line source, which we take to lie along a line in $\bR^4$ that is displaced by an amount $b$ along $x_2$ and then rotated relative to the first line source about the origin in the $x_1$-$x_3$ plane.  The harmonic forms for this second line source are again~\eqref{harmfns} in the displaced and rotated coordinates, and the geometry is then determined by the superposition of these two sources.
The fibered $y$-circle grows from zero size, reaches a maximum, and shrinks back to zero size as one moves along a path from one line source to the other, forming a homological $\bS^2$.  This $\bS^2$ has minimal size along the $x_2$ axis between the two line sources at their point of closest approach.  The $\bS^2$ shrinks to zero size as $b\to 0$, leading to an $A_1$ singularity.

The fivebrane source is actually spiralling along the T-dual of the $y$-circle and $x_1$ as in figure~\ref{fig:CircularST} (the pitch of this spiral is related to the parameter $\kappa$), and the $B$-flux through the homological $\bS^2$ is determined by the separation of the source strands along the dual geometry.  When the source strands intersect along $\bR^4$, they intersect along the dual geometry and thus the $B$-flux through the minimal $\bS^2$ vanishes.  

D3-branes wrapping this cycle have vanishing tension.  These branes are pinned to the fivebrane worldvolume; they have one remaining worldvolume direction, which wanders along the $\bT^4$; when the profile self-intersects in the five spatial dimensions of $AdS_3\times\bS^3$, such a brane becomes an effective tensionless string bound to the pair of fivebranes at their intersection.%
\footnote{More precisely, the tension is not zero, rather when $n_5$ fivebranes come together the tension is $n_5$ times smaller than that of the fundamental string.  This is the scale of the radius of curvatuce of the ambient $AdS_3\times\bS^3$ geometry, and so where deconfined, these effective strings are at their correspondence point~\rcite{Martinec:2019wzw}.}
This is the realization of a ``W-string'' of the nonabelian little string dynamics that governs coincident fivebranes.

Our picture of singularity development is thus somewhat different from that envisioned in~\rcite{Marolf:2016nwu}.  Rather than shedding angular momentum to become more compact, the system can simply wander the supertube configuration space at fixed angular momentum until it reaches a point of self-intersection of the source profile.  At this point a ``tensionless'' string singularity arises, leading to strong-coupling dynamics.
These strings can trap the fivebranes, binding them together and making a small black hole; they may also condense and split the single wrapped fivebrane into two fivebranes.

As in other examples of AdS/CFT duality, the transition to the black hole phase is one of deconfinement of non-abelian degrees of freedom in the underlying brane dynamics.  In this case, those non-abelian excitations are little strings.  Here we see the realization of this phenomenon on the {\it bulk} side of the duality.  One expects that the injection of energy above extremality will lead to a thermal gas of such W-strings which traps the pair of fivebranes and realizes a small black hole in $AdS_3\times \bS^3$.%
\footnote{Another phenomenon implemented by the condensation of these strings is the NS5 splitting transition in which the single wound fivebrane splits in two by reconnecting the intersecting strands.  This D-brane condensation is the S-dual of the corresponding topological transition of D-branes mediated by the condensation of open strings at their intersection.}

These strings are the strong-coupling version of the cycles in the weak-coupling symmetric product orbifold which describes fractionated strings oscillating along $\bT^4$, whose entropy accounts for the black hole density of states. 

One can find indirect support for this picture of the strong-coupling dynamics by asking what happens when one eliminates the $\bT^4$ from the background~\rcite{Balthazar:2021xeh,Martinec:2021vpk}.  The worldsheet sees a 6d target space $AdS_3\times\bS^3_\flat$, \ie\ one has a non-critical string background.%
\footnote{The musical ``flat'' designation indicates that spacetime supersymmetry requires that the three-sphere transverse to the fivebranes has to be squashed.}
This background is thought to arise when $n_5$ NS5-branes wrap a vanishing four-cycle in a non-compact Calabi-Yau fourfold~\rcite{Giveon:1999zm}.  The dual CFT is a deformation of the symmetric product $(\bR\times\bS^3_\flat)^N/S_N$ that describes a Fock space of fundamental strings in the decoupled fivebranes' throat.  The $AdS_3$ radius of curvature is less than the string scale, thus the system lives on the stringy side of the correspondence transition~\rcite{Horowitz:1996nw} where there are no black holes in the spectrum~- the asymptotic density of states is a Hagedorn gas of fundamental strings~\rcite{Giveon:2005mi} rather than an ensemble of BTZ black holes.  The D-branes which could potentially make little string excitations have no transverse oscillations in this case since there is no $\bT^4$ or $K3$ for them to oscillate in, and hence have little entropy.  Thus we see that when there is such a little string configuration space, there is a BTZ spectrum; when there is no such configuration space, BTZ black holes are absent from the spectrum.


\subsection{A Landau-Ginsburg dual} 
\label{sec:LGdual}

One might worry that the 1/2-BPS near-source geometry is unreliable due to possibly large $\alpha'$ corrections as a result of large curvatures there.  However, perturbative corrections in $\alpha'$ of this sort are forbiddent by the amount of symmetry in these backgrounds~-- the hyperk\"ahler nature of the transverse space geometry $ds_\perp^2$ in~\eqref{10dmetric} together with the presence of two null Killing vectors~\rcite{Horowitz:1994rf,Tseytlin:1996yb}.  

Non-perturbatively, the near-source structure is captured by a worldsheet dual description~\rcite{%
Ooguri:1995wj,
Giveon:1999px,
Giveon:2016dxe,
Martinec:2020gkv,
Halder:2022ykw} in a non-compact version of the Calabi-Yau/Landau-Ginsburg correspondence~\rcite{Martinec:1988zu,Greene:1988ut,Witten:1993yc}.  The duality of $\sltwo$ representations~\eqref{dualreps} in the circular supertube is associated to a dual description of the background in terms of the fundamental string winding condensate, represented by a $\cN=2$ supersymmetric worldsheet superpotential
\be
\label{circspotl}
{\cal W} =  \prod_{\ell=1}^{n_5} \Bigl( \sfZ \,e^{ik\sfv/n_5} - \mu_\ell\, e^{\sfX} \Bigr) 
~~,~~~~  \mu_\ell = e^{2\pi i\ell/n_5}  ~,
\ee
where $\sfv=v/R_y$.
Note that the zeroes of the superpotential spiral along the $y$ direction exactly as in figure~\ref{fig:CircularST}.
Deformations~\eqref{Vaadot} with polarizations in a single two-dimensional plane, $\alpha\dot\alpha=++$ or $- -$, have FZZ duals corresponding to deformations of this superpotential $\mu_\ell\to \mu_\ell(\sfv)$ with the twisted boundary condition $\mu_\ell(\sfv+2\pi)=e^{-2\pi ik/n_5}\mu_{\ell+k}(\sfv)$.%
\footnote{In particular, they are the lowest components of $\cN=2$ worldsheet chiral multiplets.}
The function $\mu_\ell(\sfv)$ on the $n_5$-fold cover of the $y$-circle is equivalent to $\sfF^{++}(\sfv)$~\rcite{Martinec:2020gkv}.
The deformed superpotential takes us in the direction of the more generic source exemplified by figure~\ref{fig:ThermalPrimary}.

The original non-compact CY/LG correspondence~\rcite{Ooguri:1995wj,Giveon:1999px} related a nonlinear sigma model on an $A_n$ singularity to an $\cN=(2,2)$ scalar field theory with a Liouville-like superpotential, basically~\eqref{circspotl} with $k=0$, with $\mu_\ell$ parametrizing half of the moduli (the other half are twisted chiral deformations).  In the present context, one allows the couplings $\mu_\ell$ in the superpotential to depend adiabatically on the null coordinate $\sfv$ subject to the twisted boundary condition.  

The extension of the $\mu_\ell(\sfv)$ to their $n_5$-fold covering space is the supertube profile $\sfF^{++}(\sfv)$, whose Fourier mode amplitudes are the coherent state parameters for a 1/2-BPS ground state (along the lines of~\eqref{roundST}) in which only the $|\alpha\dot\alpha\rangle_p=|\tight++\rangle_p$ and $|\tight- -\rangle_p$ modes are excited.  The superpotential zeroes code the locations of the fivebranes in their transverse space, and fivebrane intersections result when two zeroes coincide, \ie\ $\mu_\ell(\sfv)=\mu_{\ell'}(\sfv)$ for some $\sfv$ along $\bS^1_y$ and $\ell\ne\ell'$.  When this happens, a flat direction opens up in the bosonic potential $|\nabla\cW|^2$ that runs off to strong coupling~\rcite{Martinec:2020gkv}~-- the Liouville-like wall recedes, and the effective coupling at the wall grows due to the running of the dilaton in the direction of the Liouville field $\sfX$.

This dual representation of the worldsheet theory encodes effects that are non-perturbative in $\alpha'$ in the non-linear sigma model on the supertube geometry, and ensure that we have the correct picture of the degeneration of the supertube background.


\subsection{Enigmatic phases} 
\label{sec:enigmas}

Since the ``tensionless'' effective strings can engineer a splitting/joining transition of the background fivebranes, they can for instance allow the system to find more entropically favorable configurations, even on or very near the BPS bound.  It is known that below the BTZ black hole threshold there are additional highly entropic phases, see figure~\ref{fig:enigmatic}, known as {\it enigmatic phases}.  For low angular momentum $J_L < \frac{n_5n_1}2$, the entropically favored configuration consists of a zero angular momentum black hole, with the angular momentum carried by a supertube; for high angular momentum $J_L > \frac{n_5n_1}2$, a black ring is favored.  Note that spectral flow 
\be
\label{specflow}
L_0 \longrightarrow L_0 + \alpha J_L + \frac{N}{4} \alpha^2
~~,~~~~
J_L \longrightarrow J_L + \alpha N
\ee
with $\alpha=1$ relates the black-hole/supertube states with negative angular momentum to the black ring states with positive angular momentum.
The entropy in the black-hole/supertube phase is given as follows~\rcite{Bena:2011zw}.  Let the black hole and supertube have charge vectors
\be
\Gamma_{BH} = \big\{ 1,(0,0,0),(Q_5,Q_1,n_p),m  \big\}
~~,~~~~
\Gamma_{ST} = \big\{ 0,(0,0,1),(q_5,q_1,0),q_1q_5 \big\}  ~.
\ee
Here the first entry is the KKM charge, the first triplet lists the $(F1,NS5,KKM)$ dipole charges, the second triplet the monopole charges $(NS5,F1,P)$, and the last entry is the intrinsic angular momentum $2J_L$ of the object.  The total charges carried by the system are determined by the BPS conditions to be~\rcite{Bena:2011zw}
\be
\label{BHST charges}
n_5 = Q_5 + q_5
~~,~~~~
n_1 = Q_1 + q_1
~~,~~~~
2J_L = m + q_1q_5 + n_p
~~,~~~~
2J_R = q_1q_5 - n_p  ~.
\ee
The entropy carried by the black hole is given by
\be
S_{\rm BH}  = 2\pi \sqrt{Q_5Q_1n_p - m^2/4 } \equiv 2\pi\sqrt{D}  ~,
\ee
which is extremized for 
\be
n_5 q_1 = n_1 q_5 
~~,~~~~ m=0  ~,
\ee
with the value
\be
D = Q_5Q_1n_p - m^2/4 = n_5n_1n_p \bigg(1-\sqrt{\frac{q_5q_1}{n_5n_1}} \,\bigg)^2  ~,
\ee
where from~\eqref{BHST charges} we have $q_1q_5=2J_L-n_p$.  Thus we reproduce $S_{\rm\sst BH+ST}$ in~\eqref{enigma-ent}; the expression for $S_{\rm ring}$ follows from spectral flow~\eqref{specflow}.
This entropy is parametrically smaller than the symmetric product enigmatic phase entropy~\eqref{symprod enigma} (setting $k=1$ there), indicating that indeed some states have been lifted in the deformation across moduli space from the weakly-coupled CFT to the supergravity regime.

\begin{figure}[ht]
\centering
\includegraphics[width=0.6\textwidth]{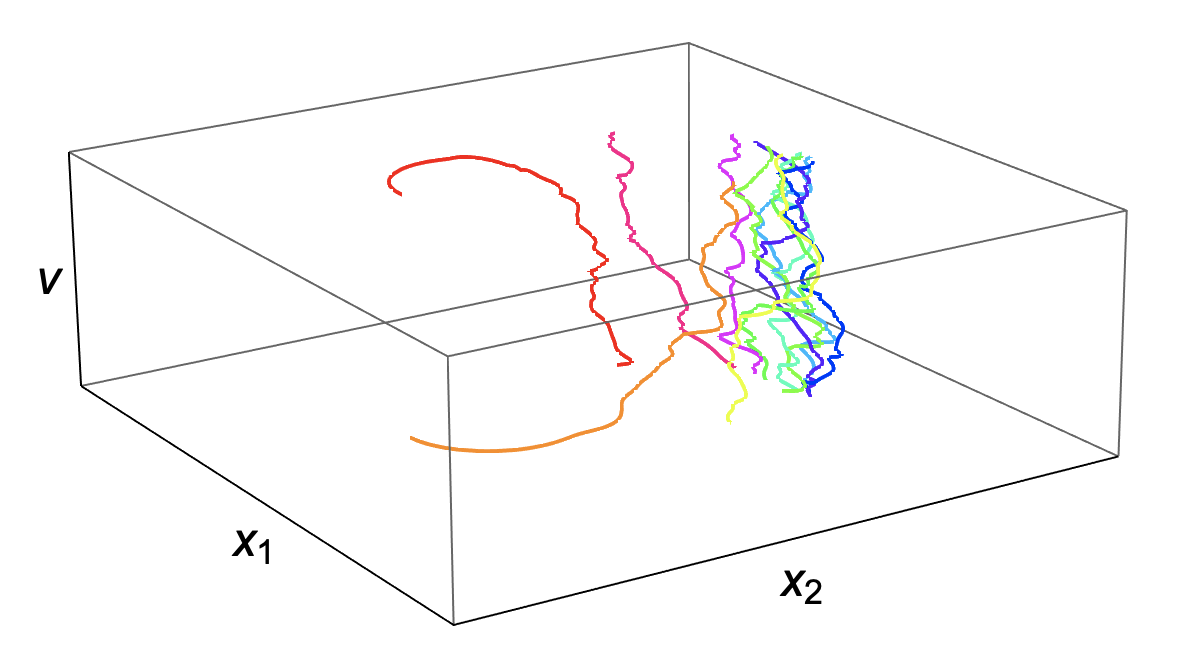}
\caption{\it A wiggly fivebrane source in a configuration where just a few fivebrane windings carry most of the angular momentum.  After exciting the system, a splitting interaction could result in two fivebranes, one of which carries most of the angular momentum, and the other most of the entropy. }
\label{fig:wigglymoulting}
\end{figure}

We can imagine this phase of two-center solutions being reached starting from an excited supertube as it wanders its configuration space and finds a point where it can split into two pieces, one of which carries away the angular momentum and the other of which carries the entropy.  See figure~\ref{fig:wigglymoulting}.  

%
%

Of course, the supertube that carries away the angular momentum need not be made out of F1 strings and NS5-branes; it could also carry F1-P charges $(q_1,q_p)$ and have zero fivebrane charge.  In other words, it can be a perturbative string carrying winding and momentum along the $y$-circle, as well as F1 dipole charge, which offloads the angular momentum of the background: 
\be
\label{qnums}
n_p = Q_p + q_p
~~,~~~~
n_1 = Q_1 + q_1
~~,~~~~
2J_L = m + \frac{q_1q_p}{d} + dn_5
~~,~~~~
2J_R = \frac{q_1q_p}{d} - dn_5  ~.
\ee
The BPS equations work the same way, with the interchange of onebrane and fivebrane quantities in the various expressions, and thus
\be
D = Q_5Q_1n_p - m^2/4 = n_5n_1n_p \bigg(1-\sqrt{\frac{q_pq_1}{n_pn_1}} \,\bigg)^2  ~.
\ee


\subsection{Elliptical training} 
\label{sec:ellipse}

An example illustrating the above structure is the elliptical deformation of a round supertube, analyzed in~\rcite{Martinec:2020gkv}.  The ellipse with semi-major and semi-minor axes $a_1,a_2$, respectively, corresponds to the symmetric product state
\be
\label{ST ellipse}
\big|\{a_1,a_2\} \big\rangle = 
\sum_{n=0}^{N/k} \Bigl[\frac{(N/k)!}{n!((N/k)-n)!}\Bigr]^{\frac 12} \Big(\frac{a_1+a_2}2\Big)^{\frac Nk-n} \Big(\frac{a_1-a_2}2\Big)^n \,
\Big(|\!+\!+\rangle_k^{~} \Big)^{\frac Nk-n}\Big(|\!-\!-\rangle_k^{~}\Big)^n
\ee
(up to normalization).

In the limit $a_2/a_1\to 0$ where the ellipse degenerates, the number of $+ +$ and $- -$ polarization modes become equal, and the angular momentum vanishes: $J_L\tight=J_R\tight=0$.  There are minimal $\bS^2$'s of the sort depicted in figure~\ref{fig:intersection} whose polar direction spans the interval between the two sides of the ellipse.  These $\bS^2$'s collapse to zero size when the ellipse degenerates.  It was indeed seen in~\rcite{Martinec:2020gkv} that the D-branes stretching across the semi-minor axis of the ellipse become massless in the DBI approximation.

Superstrata built on the elliptical supertube were constructed recently in~\rcite{Ganchev:2022exf}.  Starting with the supertube base~\eqref{ST ellipse} constructed in~\rcite{Martinec:2020gkv}, momentum was introduced via excitations $(\sfL_{-1/k})^\sfn$ on each of the $|\!-\!-\rangle_k$ cycles.%
\footnote{The parametrization of the solution in section 5 of~\rcite{Ganchev:2022exf} (after the spectral flow of section 4.4) is related to that of~\rcite{Martinec:2020gkv} as follows:
The elliptical superstratum is characterized by a set of parameters and coordinates
\be
2n+1=q_1 ~~,~~~~ \beta=\lambda q_1 ~~,~~~~
\gamma_1=\lambda^2
~~,~~~~
\gamma_2=1 ~.
\ee
The elliptical supertube then corresponds to the specialization $n=0$, with the parameters and coordinates $(a,\beta;\xi,\varphi,\chi)$ of the this supertube limit of the superstratum~\rcite{Ganchev:2022exf} related to the parameters and coordinates $(k,a_1,a_2;r,\phi,\psi)$ of the elliptical supertube solution of~\rcite{Martinec:2020gkv} via
\begin{align}
\begin{split}
k=1 
~~&,~~~~
a_1a_2=a^2 
~~,~~~~  
\frac{a_1^2-a_2^2}{2a^2} = \frac{2\beta}{1-\beta^2} 
\\[.1cm]
r^2 = -a^2F(\xi) &= \frac{a^2}{2}\left[ \frac{1+\xi^2}{1-\xi^2}-\frac{1+\lambda^2\xi^{2}}{1-\lambda^2\xi^{2}}\right]
~~,~~~~
\phi = \varphi  ~~,~~~~  \psi = \chi ~.
\end{split}
\end{align}
}
The corresponding supergravity solution was obtained by solving the hierarchy of BPS supergravity equations in a consistent truncation to three dimensions, and then lifting the solution back up to 6d.%
\footnote{More precisely, what was constructed is a spectral flowed solution that turns out to be $v$-independent and thus simpler to analyze; the unflowed background is then a large gauge transformation of this solution, described in~\rcite{Ganchev:2022exf}.  None of this changes the essential physics of the limit under discussion, namely the lengthening $AdS_2$ throat.}

The limit $a_2/a_1\to 0$ in the supertube leads to zero angular momentum, a degeneration of the geometry with collapsed cycle singularities and tensionless strings.  In the superstratum, in this same zero angular momentum limit, one finds instead that a capped $AdS_2$ throat develops and lengthens; at strictly zero angular momentum the cap descends to infinite redshift, while the eccentricity of the ellipse stays finite, puffed up by the back-reaction of the momentum wave carried by the background.  

This result sharpens a central question in the fuzzball program: Do the stringy degrees of freedom responsible for black hole entropy have a coherent wavefunction that persists out to the horizon scale of the geometry deduced from effective field theory?  Here we have seen how stringy degrees of freedom, having the same properties as those that account for the entropy at weak coupling in the CFT, arise at particular points in the space of NS5-F1 supersymmetric ground states where the geometry degenerates.   These stringy degrees of freedom are those of the nonabelian dynamics of coincident fivebranes, consistent with the idea that the Hawking-Page phase transition on the gravity side of the duality corresponds to the deconfinement transition of the gauge theory side.  The CFT is of course strongly coupled in the geometric regimes of the moduli space.

But now we also see that if we add momentum along the $y$-circle to the background~-- the third charge needed to make a BPS black hole with a macroscopic horizon~-- then in the same limit that exhibited tensionless strings in the two-charge NS5-F1 solution, in the three-charge NS5-F1-P supergravity solution an extremal $AdS_2$ black hole throat develops.  Is this horizon in the three-charge states a manifestation of the same tensionless strings that appear in the two-charge states?  How does one reconcile this with the apparent smoothness of the horizon in the three-charge geometry?
A goal of future work will be to look for evidence of tensionless strings at the horizon of this three-charge geometry, which would indicate that the horizon seen in the bulk effective field theory is not actually a horizon for the fundamental black hole constituents.


\section*{Acknowledgements}

We thank 
Davide Bufalini, 
Sergio Iguri,
and
Nicolas Kovensky
for discussions.
EJM thanks the organizers of Strings 2022 in Vienna for their hospitality, and the invitation to present our results. We also thank the IPhT, CEA Saclay for hospitality during the course of this work.
This work is supported in part by DOE grant DE-SC0009924.
The work of DT was supported by a Royal Society Tata University Research Fellowship. The work of SM is supported by the MIUR program for young researchers ``Rita Levi Montalcini''.


\appendix


\section{Conventions} 
\label{sec:conventions}

In this paper we use conventions largely in parallel with those of~\cite{Martinec:2020gkv}, with some differences that we record below. Denoting quantities in~\cite{Martinec:2020gkv} with tildes, and those of the present work without tildes,
the conventions for the Cartan angles of the target-space $\mathbb{S}^3$ are related by
\be
\label{eq:flipang}
\psi=\tilde\phi \;, \quad~~
\phi=\tilde\psi \;, \quad~~
\theta=\frac{\pi}{2}-\tilde\theta
\ee
which results for instance in $J^3_\su=-{\tilde J}{}^3_\su$. Correspondingly we have set $l_2=-1$ in \eqref{nullcoeffs}, where $\tilde{l}_2=1$ was used in~\cite{Martinec:2020gkv}. 
We have chosen this convention in order to work with lowest-weight states in SU(2), see e.g.~equation~\eqref{halfBPS}. The map to the conventions of~\cite{Bufalini:2021ndn} is to perform~\eqref{eq:flipang}, then to send either $t \to -t$ or $(y,\phi,\psi)\to (-y,-\phi,-\psi)$, and finally to send $\mathsf{k}\to -\mathsf{k}$ (which here is the parameter $k$).

The NS5-F1 circular supertube supergravity fields, in the fivebrane decoupling limit, in our conventions are 
\begin{align}
\label{NS5F1}
ds^2 &= \Bigl( -du\:\! dv + ds_{\scriptscriptstyle\mathbf T^4}^2  \Bigr)
+ {n_5}\Bigl[ d\rho^2+ d\theta^2+\frac1\Sigma \Bigl( {\cosh}^2\!\rho\sin^2\!\theta \,d\phi^2 + {\sinh}^2\!\rho\cos^2\!\theta \,d\psi^2 \Bigr)\Bigr] 
\nonumber\\[.1cm]
& \hskip .6cm 
- \frac{2\alphab}{\Sigma} \Bigl(  {\sin^2\!\theta \, dt\,d\phi + \cos^2\!\theta \, dy\,d\psi}  \Bigr)
+\frac{\alphab^2}{\nfive\Sigma} \Bigl[ \nfive \sin^2\!\theta \, d\phi^2 +  \nfive\cos^2\!\theta \, d\psi^2  
+ du\:\! dv \Bigr],
\nonumber\\[8pt]
B  &= \frac{ \cos^2\!\theta (\alphab^2+\nfive\cosh^2\!\rho)}{\Sigma}  { d\phi\wedge d\psi - \frac{\alphab^2}{n_5\Sigma} \, dt\wedge dy } \nonumber\\
& \hskip .6cm 
{}-\frac{\alphab  \cos^2\!\theta}{\Sigma}  dt\wedge d\psi
- \frac{\alphab  \sin^2\!\theta}{\Sigma}  dy\wedge d\phi~, 
\qquad\qquad\quad u=t+y \;, ~~v=t-y \;,
\nn\\[8pt]
e^{-2\Phi} & = \frac{n_1\Sigma}{k^2\Ry^2\,V_4} ~,\qquad~~~  
\Sigma = \frac{\alphab^2}{\nfive} + \sinh^2\rhoo + \cos^2\theta ~, \qquad\quad \alphab\equiv k\Ry \;.
\end{align} 
The periodic identification of $y$, namely $y \sim y+2\pi \Ry$, induces a local $\bZ_k$ orbifold singularity at the supertube location $\rho=0$, $\theta=\pi/2$. The gravitational angular momenta are 
\begin{align}
\begin{aligned}
J^3&\;=\;\frac{1}{2} (J^{\phi}-J^{\psi})
\;=\;\frac{1}{2}\frac{n_1 n_5}{ k }\,,
\qquad~~~
\bar J^3&\;=\;\frac{1}{2} (J^{\phi}+J^{\psi})
\;=\;\frac{1}{2}\frac{n_1 n_5}{ k }\;.
\label{gravangmom}
\end{aligned}
\end{align}
After taking the $AdS_3\times \mathbb{S}^3$ limit, the holographically dual state of the symmetric product orbifold CFT is $(\ket{++}_k)^{N/k}$.


\section{Supertube geometry} 
\label{sec:STgeom}

The ten-dimensional geometry sourced by the NS5-F1 supertube source is given by~\rcite{Skenderis:2006ah,Kanitscheider:2006zf,Kanitscheider:2007wq,Giusto:2013rxa,Bena:2015bea,Giusto:2015dfa} (see also \rcite{Bakhshaei:2018vux})
\begin{subequations}
\label{STfields}
\allowdisplaybreaks
\begin{align}
d s_{10}^2 &= 
-\frac{Z_5}{\cP}\Bigl[\big(du\!\!\,+\!\omega \big) \big(dv\!\!\,+\!\beta\big) \Bigr] 
+ Z_5 \,ds_{\!\perp}^2 +  \,ds_{\cM}^2
 ~,  \label{10dmetric}
\\[.2cm]
e^{2\Phi}&=g_s^2\frac{Z_5^2}{\cP}
\\[.2cm]
B_2 &= \frac{Z_5}{2\cP}\,(d u +\omega) \wedge(d v  +  \beta) + 
\sfb_{ij} \,dx^i\wedge dx^j  
\\[.2cm]
\label{C0}
C_0&= -\frac{Z_0}{Z_5} 
\\[.2cm]
\label{C2}
C_2&= +\frac{Z_{(\gamma)}}{Z_5}\Omega^{(\gamma)} 
+ \frac{Z_0}{2\cP}\,(d u+\omega) \wedge(d v+\beta)  
+\sfc_{ij} \,dx^i\wedge dx^j 
\\[.2cm]
\label{C4}
C_4 &= -\frac{Z_0}{Z_5}\, \widehat{\mathrm{vol}}_{4}  
- \frac{Z_0}{2\cP}\Big[ \big(du+\omega\big)\wedge \big(dv+\beta\big)\wedge \big(\sfb+\half \omega\wedge\beta\big) - \omega\wedge\beta\wedge \sfb \Big]
\\
&\hskip 1.5cm
+\half\,\big(du+\omega\big)\wedge \big(dv+\beta\big)\wedge\Big( \sfc + \frac{Z_{(\gamma)}}{Z_5}\, \Omega^{(\gamma)} \Big) 
-\Big( \sfc_{(\gamma)} + \frac{Z_{(\gamma)}}{Z_5} \,\sfb\Big)\wedge \Omega^{(\gamma)}
\nn
\end{align}
\end{subequations}
with
\begin{equation}
\cP   \equiv     Z_1 \, Z_5  - Z_0^{\,2} -  Z_{(\gamma)}^{\,2} ~.
\label{curlyP again}
\end{equation}
Here $ds^2_{10}$ is the ten-dimensional string-frame metric, $ds_\perp^2$ is the metric on the space transverse to the branes, $\Phi$ is the dilaton, $B_p$ and $C_p$ are the NS-NS and R-R gauge forms. The volume form on $\cM$ is denoted by $\widehat{\mathrm{vol}}_{4}$, and $\Omega^{(\gamma)}$ is a basis of anti-selfdual two-forms on $\cM$ (so 3 for $\bT^4$, 19 for $K3$). 
The various harmonic forms and functions are related by
\begin{align}
d\sfb = \ast_\perp dZ_5
~~,~~~~
d\sfc = \ast_\perp dZ_0 
~~,~~~~
d\sfc_{(\gamma)} = \ast_\perp dZ_{(\gamma)}
\end{align}
also, $\omega$ is self-dual while $\beta$ is anti-selfdual in the transverse $\bR^4$ parametrized by $x^i$.

The R-R fields are odd under $Z_0\to -Z_0,Z_{(\gamma)}\to - Z_{(\gamma)}$, while the NS-NS fields are even.  On $\bT^4$, these scalar and anti-selfdual tensor coefficient functions are related to those in~\eqref{LMints} by
\be
Z_0 = Z_{[AB]}
~~,~~~~
\big\{ Z_{(\gamma)} \big\} = \big\{ Z_{(AB)} \big\} ~,
\ee
and correspond to the ``internal'' excitations of the supertube.

The bosonic field content of $\cN=(2,2)$ 6d supergravity obtained upon dimensional reduction on $\cM=\bT^4$ consists of the graviton, 10 tensors (5 SD and 5 ASD, one each NS-NS and the rest R-R), 16 vector multiplets (8 NS-NS and 8 R-R), and 25 scalars.  Of the latter, 5 are fixed scalars and the remaining 20 parametrize the moduli space $\big(\frac{O(5,4)}{O(5)\times O(4)}\big)/\Gamma_\sfq$. 

The content of the $\cN=(2,0)$ 6d supergravity for $\cM=K3$ consists of the supergravity multiplet (the metric plus 5 SD tensors) together with 21 ASD tensor multiplets (each with 1 ASD tensor and 5 scalars).  The moduli space is $\big(\frac{O(5,21)}{O(5)\times O(21}\big)/\Gamma_\sfq$, and again there are 5 fixed scalars.


\section{Spacetime supersymmetries} 
\label{sec:BPScondition}

The spacetime supersymmetry charges take the worldsheet form~\rcite{Giveon:1998ns,Giveon:1999jg,Berenstein:1999gj,Martinec:2020gkv}
\be
\label{susyQ}
\sfQ_{\!\vec{\,\vareps}} = \oint \!dz\, e^{-(\varphi-\tilde\varphi)/2} \, S_{\!\vec{\,\vareps}}
\quad,\qquad
S_{\!\vec{\,\vareps}} = \exp\bigg(\frac i2\sum_{i=1}^6 \vareps_i H_i\bigg) ~.
\ee
These operators are then subject to the GSO projection and the BRST constraints.

The usual type IIB GSO projection in the worldsheet theory on global $AdS_3\times\bS^3$~\rcite{Giveon:1998ns} sets
\be
\label{10dGSO}
\prod_{i=1}^5 \vareps_i = 1 ~~,~~~~
\prod_{i=1}^5 \bar\vareps_i = 1 ~,
\ee
hence in the null-gauged WZW model we should have the same requirement, at least to leading order in $1/R_y$.  
In this $AdS_3$ decoupling limit, the three-fermion term in the $\gamma G$ BRST constraint then enforces
\be
\label{Gconstraint}
\vareps \equiv \vareps_1\vareps_2\vareps_3 = -1 ~,
\ee
and similarly $\bar\vareps\equiv \bar\vareps_1\bar\vareps_2\bar\vareps_3=-1$.
The null currents $\cJ,\bar\cJ$ constrain
\be
\label{Jconstraint}
\vareps_1+\vareps_2 l_2 = 0
~~,~~~~
\bar\vareps_1+\bar\vareps_2 r_2 = 0 ~.
\ee
The superpartners of the null currents are given by
\begin{align}
\begin{split}
\lambdab &\,=\, \sqrt{n_5}\big(\psi^3_\sl+l_2\psi^3_\su\big)+l_3\psi^t+l_4\psi^y
\,=\, -\sqrt{n_5}\,e^{-iH_3} +kR_y\, e^{-iH_6}\;,
\\[.2cm]
\bar\lambdab &\,=\, \sqrt{n_5}\big(\bar\psi^3_\sl+r_2\bar\psi^3_\su\big)+r_3\bar\psi^t+r_4\bar\psi^y
\,=\, -\sqrt{n_5}\,e^{-i\bar H_3} - kR_y\, e^{i\bar H_6}  ~,
\end{split}
\end{align}
where we have used the NS5-F1 supertube null vector coefficients~\eqref{nullcoeffs} and the bosonization formulae~\eqref{eq:bosonizations}.

The null BRST supercurrents $\tilde\gamma\lambdab,\overline{\tilde\gamma}\bar\lambdab$ constrain linear combinations $c_{\!\vec{\,\vareps}}S_{\!\vec{\,\vareps}}$.
Denoting the coefficients $c_{\vareps_3\vareps_6}$ and suppressing the labels $\vareps_1,\vareps_2,\vareps_4,\vareps_5$ (which we hold fixed), the allowed nonzero coefficients are 
\begin{align}
\begin{split}
\label{lamconstraint}
c_{--} ~~,~~~~ c_{-+} = \frac{\sqrt{n_5}}{kR_y} c_{+-}
\quad;\qquad
\bar c_{-+} ~~,~~~~ \bar c_{--} = -\frac{\sqrt{n_5}}{kR_y} \bar c_{++} ~,
\end{split}
\end{align}
in particular $\vareps_6=-\bar\vareps_6$ for the corresponding solutions.  Only the first solution in each left/right chirality is compatible with~\eqref{Gconstraint}, \eqref{Jconstraint}, which set $\vareps_3=\bar\vareps_3=-1$ and so $c_{+-}=\bar c_{++}=0$.  Note that in order to impose the 10d GSO projection~\eqref{10dGSO}, we must have {\it opposite} 12d GSO projections on left and right.  From \eqref{eq:GSO-L},  \eqref{eq:GSO-R} we have
\be
\label{GSOproj}
\prod_{i=1}^6 \vareps_i = -1 ~~,~~~~
\prod_{i=1}^6 \bar\vareps_i = +1 ~,
\ee
Overall, the physical supercharges~\eqref{susyQ} are labeled by (at leading order in $1/R_y$)
\begin{align}
\begin{split}
\label{Qdynkin}
\vareps_1 = \vareps_2 \equiv \alpha
~~,~~~~
\vareps_3 = -1 
~~&,~~~~
\vareps_6 = -1
~~,~~~~
\vareps_4 = -\vareps_5 \equiv \dot A
\\[.2cm]
\bar\vareps_1 = \bar\vareps_2 \equiv \dot\alpha
~~,~~~~
\bar\vareps_3 = -1 
~~&,~~~~
\bar\vareps_6 = +1
~~,~~~~
\bar\vareps_4 = -\bar\vareps_5 \equiv \dot B
\end{split}
\end{align}

Spectral flow in the spacetime R-charge by an amount $\delta$ shifts the supercurrent modings via
\be
\sfG_n^{\alpha \dot A} \longrightarrow \sfG^{\alpha\dot A}_{n-\alpha\delta} ~.
\ee
The supertubes correspond to Ramond ground states with R-charge between $N/2$ and $N$, which in the NS sector are antichiral states with R-charge between $-N/2$ and $0$.  These antichiral states are annihilated by $\sfG_{-1/2}^{-\dot A}$ and $\sfG_{+1/2}^{+\dot A}$, and so the corresponding supertube states are annihilated by
\be
\sfG_{0}^{-\dot A}
~~,~~~~
\sfG_{0}^{+\dot A}
~~,~~~~
\bar\sfG_{0}^{-\dot B}
~~,~~~~
\bar\sfG_{0}^{+\dot B}
~~.
\ee
These are precisely the global supercharge operators~\eqref{susyQ} with the polarization states~\eqref{Qdynkin}.%
\footnote{Note that the $\sltwo$ and $\sutwo$ polarizations reflect the quantum numbers before spectral flow, $n=\half\vareps_1,\alpha=\vareps_2$.}

The bosonized ghost exponential $e^{-\varphi/2}$ of the supercharge~\eqref{susyQ} has a square root singularity with respect to the $(-1)$ picture NS-NS vertex operators~\eqref{NSops}.  For a vertex operator to commute with the supercharge, the fermion $\psi$ should have a square root zero $\psi(z) S(w) \sim \sqrt{z-w}$.  Similarly, for the $(-1/2)$ picture R-R operators in the specified null superghost pictures ($(+1/2)$ for both the supercharge and the vertex operator), one gets a bosonized ghost OPE singularity $(z-w)^{-1/2}$, requiring a spin field OPE $(z-w)^{+1/2}$.  

Vertex operators that preserve the BPS property should commute with these supercharges.  In~\eqref{NSops}, the $\sutwo$ fermion polarization in $\cX$ should be $\psi^-_\su$ in order to be BPS (this is the operator $\cX^+$ for which there is only one term in the Clebsch); then the fermion polarization is such that the OPE with the spin field scales as $\sqrt{z-w}$, and the vertex operator commutes with the supercharges.
Similarly, the Clebsch of the $\sltwo$ fermion in $\cW^-$ guarantees that it also commutes with the supercharges.
These two polarization choices correspond to the operators $\cV^\pm$ defined in~\eqref{halfBPS}.  

Similarly, the spin field polarizations in~\eqref{Rops} that commute with the above supercharges are the $\cS^A$ of~\eqref{halfBPS}.  Adding up the spin field contribution $\!\vec{\,\vareps}_Q^{~}\cdot\!\vec{\,\vareps}_\cS^{~}$ to the OPE singularity, one sees that indeed the spin field OPE contributes $\sqrt{z-w}$ and indeed the supercharges commute with $\cS^A$.

One can further check that the other vertex operators we have described above do not commute with some of these supercharges.


\section{The shockwave limit} 
\label{sec:shockwave}

\subsection{The Lunin-Mathur source integrals}
\label{sec:LMints}

Our analysis follows that in~\rcite{Lunin:2002iz,Bena:2016agb}.
The circular supertube profile is given by
\be
\label{circular}
\sfF^{++} 
= a_k\, \exp[2\pi i k \hat v/L] ~.
\ee
where $\hat v\in [0,L]$, $L=2\pi n_5/R_y$ parametrizes the $n_5$-fold covering space of the $y$-circle in the NS5-P duality frame.
It will prove convenient to denote 
$\sfx=x^{++}$, 
$\sfy=x^{+-}$, 
and parametrize the profile by $\xi\equiv2\pi k\hat v/L$. Since the supertubes of interest run around the same profile $k$ times, the integral is simply $k$ times the integral over the range $\xi\in[0,2\pi)$.
The further change of variables $z=e^{i\xi}$, and the use of $\bar z =1/z$ for an integral along the unit circle in $z$, converts the integrals into contour integrals for which we can use the method of residues, for example
\be
\label{Hint}
Z_5=\frac{Q_5}{2\pi i}\oint \frac{dz}{z} \frac{1}{(\sfx-a z)(\bar \sfx - a/z)+\sfy\bar \sfy} = \frac{Q_5}{\sqrt{\tilde{w}^2-4\sfx\bar \sfx a^2}}  ~.
\ee 
where $\tilde{w}=\sfx\bar \sfx+\sfy\bar \sfy+a^2$.
Converting from Cartesian coordinates to spherical bipolar ones
\begin{align}
\sfx = \sqrt{a^2+r^2}\,\sin\theta \,e^{i\phi}
~~&,\quad
\sfy =  r \cos\theta \,e^{i\psi} 
\end{align}
leads to
\be
\label{Hanswer}
Z_5 = \frac{Q_5}{r^2+a^2\cos^2\theta} =\frac{Q_5}\Sigma ~.
\ee
Next, we introduce an $\sfF_0$ term to the profile function,
\be
\sfF_0(\hat v) ~=~ \epsilon^{AB}F_{AB}(\hat v)
~=~ - \frac{2b_0}{\nu k R_y} \sin \left( \frac{2\pi k}{L}  \nu \, \hat v \right) ~=~ \frac{-b_0}{i\nu k\R} \, (z^\nu-z^{-\nu})  \,,
\label{g5}
\ee
where $b_0$ is real. 
The $\sfF_0$ term in the profile function gives rise to the following contour integral expression for the harmonic function $Z_0$:
\be
\label{Aint}
Z_0 = \frac{b_0}{2\pi i}\oint \frac{dz}{z} \frac{z^\nu+z^{-\nu}}{(\sfx-a z)(\bar \sfx-a/z)+\sfy\bar \sfy} 
~=~ 2 b_0 \left(\frac{a^2 \sin^2\theta}{r^2+a^2}\right)^{\nu/2} \frac{\cos \nu\phi}{\Sigma} \,.
\ee
One also has the fibration one-form, $\beta$, given by 
\begin{equation}
\beta ~=~  \frac{R_y \, a^2}{\sqrt{2}\,\Sigma}\,(\, \sin^2\theta\, d\phi - \cos^2\theta\,d\psi\,)   \,,
 \label{betadefn}
\end{equation}
and the angular momentum one-form
\begin{equation}
\omega ~=~ \omega_0 \,, \qquad \omega_0 ~\equiv~  \frac{a^2 \, R_y \, }{ \sqrt{2}\,\Sigma}\,  (\sin^2 \theta  \, d \phi + \cos^2 \theta \,  d \psi ) \,.
\label{angmom0}
\end{equation}

The shockwave limit takes $\nu$ to be extremely large, so that the profile~\eqref{Aint} localizes around the circular supertube source at $r=0,\theta=\frac\pi2$.  Note that not only are the R-R fields~\eqref{C0}-\eqref{C4} highly localized as a result, they are also rapidly oscillating along this circle, so that on average the R-R fields are essentially invisible.

The source function $\sfF_0$ also contributes to the one-brane harmonic function $Z_1$
\be
\label{F5dot}
|\dot \sfF_0|^2 = \Big(\frac{2b_0}{kR_y}\Big)^2 \cdot \half\Big[1+\cos\Big(\frac{4\pi k}{L}\,\nu\hat v\Big)\Big] ~.
\ee
The cosine term leads to another highly localized, rapidly oscillating contribution that is invisible in the large $\nu$ limit.  The constant term contributes a term identical to the $|\dot \sfF_{\alpha\dot\alpha}|^2$ contribution, leading to a relation between the scales $a,b_0$ and the charges $Q_1,Q_5$
\be
\label{windbudget}
a^2+\frac{2|b_0|^2}{k^2 R_y^2} \,=\, \frac{Q_1Q_5}{k^2R_y^2}~.
\ee
This equation is simply the F1 winding budget $kn^{++}_k + (\nu k) n^{00}_{\nu k} = n_1n_5$, expressed in terms of coherent state parameters.

The onebrane charge is given by
\begin{equation}
\label{Q1int}
 Q_1=\frac{Q_5}{L}\int_0^L \bigl(|\dot{\sfF}_{\alpha\dot\alpha}(\hat v)|^2+|\dot{\sfF}_{AB}(\hat v)|^2\bigr)d\hat v ~.
\end{equation}
The quantities $Q_1$, $Q_5$ are related to quantized onebrane and fivebrane
numbers $n_1$, $n_5$ by
\begin{equation}
Q_1 = \frac{n_1\,g_s^2\,\alpha'^3}{v_4}\,,\qquad Q_5 = n_5\,\alpha' ~,
\label{Q1Q5-n1n5}
\end{equation}
where $v_4$ is the volume of $\bbT^4$ in string units. 

When $b_0\ne0$, the coefficients of the harmonic function are altered, in that the radius $a$ of the supertube is reduced due to the devotion of some of the winding budget~\eqref{windbudget} to cycles of type $\epsilon^{AB}\ket{AB}_{\nu k}$ (also called ``$\ket{00}$'' cycles in the literature); this changes the coefficient of $Z_5$ relative to $Z_1$.  As a result, the metric no longer has the tuning of coefficients that allows the supertube locus at $r=0,\theta=\frac\pi2$ to be nonsingular, and instead one finds a shockwave singularity there~\cite{Lunin:2002bj,Lunin:2002iz}.  Restoring the rapidly oscillating terms in~\eqref{F5dot}, which according to~\eqref{Aint} only contribute close to the source locus, smooths out the singularity.

Note that if we had not a single mode but a distribution of high frequency modes, then in $|\dot \sfF_0|^2$ two high frequency modes can generate a low but non-zero frequency in their product which would lead to a low-frequency modulation of $Z_1$.  In order to avoid this, and just have a shockwave, we can for instance have all the high frequency modes be multiples of some large $\kappa\gg1$, with $k_i\propto \kappa$ and then $|\dot \sfF_0|^2$ has a constant frequency piece and then higher modes of frequency at least $\kappa$.


\subsection{Shockwaves in 3-charge backgrounds}
\label{sec:GLMTshocks}

The analysis of trapping behavior in~\rcite{Eperon:2016cdd} included a class of NS5-F1-P 3-charge geometries obtained by ``fractional spectral flow'' of the 1/2-BPS ground states~\eqref{roundST},~\cite{Lunin:2004uu,Giusto:2004id,Giusto:2004ip,Giusto:2012yz,Chakrabarty:2015foa}.  Shockwaves in these geometries were considered in~\rcite{Chakrabarty:2021sff}.  These backgrounds also have an exact worldsheet description as null-gauged WZW models~\rcite{Martinec:2017ztd,Martinec:2018nco,Martinec:2019wzw}, and as in section~\ref{sec:stability}, one can again ask where the vertex operator wavefunctions localize in the large $j'$ limit, and make a shockwave that is regularized by stringy effects.

The fractional $\sutwo$ spectral flow of the state by an amount $s/k$ makes an allowed state in the spacetime CFT~\rcite{Giusto:2012yz}, see also~\rcite{Chakrabarty:2015foa}.  In the worldsheet description, the null vector coefficients are modified to
\begin{align}
\begin{split}
l_2 = -(2s+1)
~~,~~~~
r_2=-1
~~,~~~~
l_3 &= r_3 =  -\bigg( kR_y+\frac{n_5 s(s+1)}{kR_y}\bigg)
\\[.3cm]
l_4 = kR_y + \frac{n_5 s(s+1)}{kR_y}
~~,~~~~
r_4 &= -kR_y + \frac{n_5 s(s+1)}{kR_y} ~.
\end{split}
\end{align}
The geometry has an evanescent ergosurface (a quadratic vanishing of the metric coefficient $g_{uv}$; or geometrically, the surface on which the Killing vectors $\partial_y$ and $\partial_u$ are orthogonal) on the locus
\be
\label{evanergo}
r=0
~~,~~~~
\tan^2\theta=\frac{s+1}s  ~.
\ee

The vertex operators must satisfy the null constraints $\cJ=\bar\cJ=0$ with these modified coefficients in the null vectors.  These constraints were analyzed in~\rcite{Martinec:2018nco}.  Supergravity vertex operators again have $w_y=0$.  The axial null constraint imposes
\be
\label{GLMTnull}
m - \bar m = -kn_y + (2s+1)m' - \bar m' 
\ee
up to shifts of order one.

Let us look for vertex operators concentrated on the evanescent ergosurface~\eqref{evanergo}.  The 1/4-BPS center-of-mass wavefunctions are
\begin{align}
\begin{split}
\Phi_{j;-j-\sfn,-j}^{\sl} &= 
e^{ i(2j+\sfn)\tau - i\sfn\sigma }
\, r^\sfn \,\Big(\frac{a^2}{r^2+a^2}\Big)^{j+\sfn/2} + O(1/j)
\\[.3cm]
\Psi_{j';-j'+\sfm,-j'}^{\su} &=
e^{ i(2j'-\sfm) \phi + i\sfm \psi }
\big(\sin\theta\big)^{2j'-\sfm} \big(\cos\theta\big)^{\sfm} 
\end{split}
\end{align}
The null constraint~\eqref{GLMTnull} is solved for large $j'$ by \be
m\sim \bar m\sim j
~~,~~~~
n_y\sim 0
~~,~~~~
m' \sim \frac{j'}{2s+1}
~~,~~~~
\bar m'\sim j' ~~,
\ee
for which the peak is at~\eqref{evanergo}.  For large $j'$ (and thus large $n_5$), the string wavefunction is concentrated within a string scale distance of the evanescent ergosurface, just as we saw for the 1/2-BPS two-charge geometries.


\vskip 2cm

\bibliographystyle{JHEP}      

\bibliography{microstates}


\end{document}